\documentclass[aps,pre,twocolumn,superscriptaddress,amsmath,amssymb,showpacs,longbibliography]{revtex4-1}
\usepackage{graphicx}
\usepackage{color}
\renewcommand{\phi}{\varphi}

\begin{document}

\title{Hierarchical Landscape of Hard Disk Glasses}

\date{\today}

\author{Qinyi Liao}

\affiliation{CAS Key Laboratory of Soft Matter Chemistry, Hefei National Laboratory for Physical Sciences at the Microscale, and Department of Physics, University of Science and Technology of China, Hefei 230026, People's Republic of China}

\affiliation{Laboratoire Charles Coulomb (L2C), University of Montpellier, CNRS, Montpellier 34095, France}

\author{Ludovic Berthier}

\affiliation{Laboratoire Charles Coulomb (L2C), University of Montpellier, CNRS, Montpellier 34095, France}

\begin{abstract}
We numerically analyse the landscape governing the evolution of the vibrational dynamics of hard disk glasses as the density increases towards jamming. We find that the dynamics becomes slow, spatially correlated, and starts to display aging dynamics across an avoided Gardner transition, with a phenomenology that resembles three dimensional observations. We carefully analyse the behaviour of single glass samples, and find that the emergence of aging dynamics is controlled by the apparition of a complex organisation of the landscape that splits into a remarkable hierarchy of minima as jamming is approached. Our results show that the mean-field prediction of a Gardner phase characterized by an ultrametric structure of the landscape provides a useful description of finite dimensional systems, even when the Gardner transition is avoided.
\end{abstract}

\maketitle

\section{Introduction}

The Gardner transition has a long history in the spin glass literature. It was first discovered in the context of mean-field spin glass models characterized by a random first order transition, where the glassy phase was found to destabilise at low temperature into a more conventional spin glass phase~\cite{sgbook1987, GKS85, gardner1985}, characterized by a complex hierarchical organisation of pure states. The existence of a Gardner phase was more recently also proven for particle models in the absence of quenched disorder. In the mean-field limit, dense fluids first undergo a random first order transition to a glassy phase that may also transform in a marginal spin glass phase~\cite{ckpuz2014, review2010, review2011, review2016}. The existence of a Gardner transition in finite dimensional glass and spin glass models is still a debated issue, that neither theory nor simulations have been able to resolve~\cite{sgreview2000, rg2011, rg2015, rg2017, rg2018, sg2013, sg2014}.

This debate is particularly relevant in the context of hard sphere glasses, where the mean-field theory has been more precisely developed~\cite{review2010, review2016}. When the density is increased, the hard sphere fluid first undergoes a glass transition to an arrested glassy state~\cite{review2011}. Upon further compression, the hard sphere pressure increases very fast and diverges at a jamming transition. Jamming itself is a critical phenomenon, whose properties can be obtained in terms of marginal stability~\cite{jc2010, jjc2015}. In the context of mean-field theory, a marginally stable Gardner phase is encountered between the glass and jamming transitions~\cite{ckpuz2014exact, ckpuz2014, review2016}. The marginal properties of the Gardner phase are thus key to properly describe the jamming criticality within mean-field theory. This is a very puzzling observation, as the critical exponents of jamming seem to remain roughly unchanged between $d=\infty$ and $d=2$~\cite{jc2015, jjc2015}, whereas instead the Gardner transition appears fragile against finite dimensional fluctuations~\cite{rg2015, rg2017, rg2018, sz2018, 1d2018}.

These two sets of observations are hard to reconcile. On the one hand, the precise coincidence between mean-field predictions for jamming and numerical measurements in finite $d$ would suggest that there exists a large pressure regime where mean-field predictions must become accurate~\cite{ckpuz2014, review2016}. On the other hand, it is unlikely that a Gardner phase can exist in dimensions as low as $d=2$, where jamming criticality remains similar to the one in $d=\infty$~\cite{jc2009, jc2010, jc2015, jjc2015}, suggesting that the Gardner transition, in itself, is not needed for a theory to capture the jamming criticality. Resolving this paradox is important, as mean-field theory can be used to tackle a large number of physical questions from a fully microscopic perspective, such as thermodynamic properties~\cite{experiment2016, experiment2018, bcjpsz2016, ckpuz2014, sz2018}, the structure of phase space, the evolution of vibrational dynamics~\cite{modes3, modes4}, or the rheology of hard sphere glasses~\cite{shear2015, shear2017, sshear2017}. In addition, it is also important to understand under which conditions a complex free energy landscape may become physically relevant, in order to make novel predictions for experimental work dealing with the glassy dynamics of granular, colloidal, and molecular systems.

To tackle these questions we study the properties of hard disk glasses in the regime between glass and jamming transitions. For this system, it is known that jamming has the same critical properties as in any larger dimension up to $d=\infty$. It is also certain that a sharp Gardner transition cannot exist~\cite{rg2015, rg2017, rg2018}, although numerical simulations in $d=2$ spin glasses indicate that the spin glass phenomenology can indeed be observed~\cite{sg2018, 2dsg2018}. This suggests that a relatively sharp crossover should exist between two types of glass states, a simple and a marginal glass (see Refs.~\cite{bcjpsz2016,experiment2016} for earlier hints for $2d$ hard disks systems), associated to the standard spin glass phenomenology and an emerging complexity of phase space, but that this phenomenology is not the result of a phase transition.

We find that hard disk glasses indeed display the same phenomenology as three dimensional hard spheres, associated with growing (but non-diverging) timescales and lengthscales, as they are compressed towards jamming~\cite{bcjpsz2016, sz2018}. By investigating the properties of single glass samples, we find that the emergence of collective vibrational dynamics corresponds to the development of a complex organisation of the landscape and a hierarchy of barriers and lengthscales that are strongly reminiscent of the mean-field description of spin glass phases~\cite{sgbook1987,gardner1985,ckpuz2014exact}. From our results, it is not possible to decide whether the complexity of the free-energy landscape near jamming stems from the marginality of jammed states, or, rather, whether the existence of an avoided Gardner phase transition is responsible for the original properties of the jamming transition. Our study suggests that it is not possible to observe jamming criticality without the Gardner physics. To our knowledge there is also no evidence that a finite dimensional particle system exhibits marginal stability far from a jamming transition~\cite{sbz2017}, contrary to predictions obtained within mean-field theory~\cite{jg2018}.

This paper is organized as follows.
In Sec.~\ref{details}, we introduce the model, simulation methods, and define the main observables used in this work.
We present the numerical evidence for an avoided Gardner transition in Sec.~\ref{evidence}.
In Sec.~\ref{single}, we present a detailed analysis of the free energy landscape and the non-equilibrium dynamics observed in individual glass samples.
In Sec.~\ref{hierarchical}, we demonstrate the emergence of a hierarchical landscape and show that the location of the Gardner avoided transition is strongly dependent on the chosen nonequilibrium protocol.
In Sec.~\ref{larger}, we present results for larger systems and a careful analysis of finite-size effects. We conclude the paper in Sec.~\ref{conclusion}.

\section{Simulation details}

\label{details}

\subsection{Model and methods}

We simulate dense assemblies of $N$ hard disks in a $d=2$ square box with periodic boundary conditions. To avoid crystallization, polydisperse mixtures are adopted, where the diameters of the disks are drawn from the continuous distribution $P(\sigma) \propto \sigma^{-3}$, with $\sigma_{min} / \sigma_{max} = 0.45$, $\sigma_{min}$ and $\sigma_{max}$ are the minimum and the maximum diameters, respectively. The simulation is carried out using Monte Carlo (MC) simulations, using two types of algorithms. Throughout this work, the time unit is defined as a MC sweep, i.e. a series of $N$ attempts to perform a translational move for a single particle chosen at random.
The units of length and energy are set by the average diameter $\bar{\sigma} = \frac{1}{N} \sum_{i} \sigma_{i}$ and temperature $k_{B}T$, respectively.
The pressure is reduced by the factor $k_{B}T \rho$, where $\rho$ is the number density. The dimensionless pressure $Z = P /(k_B T \rho)$ is extracted from the rescaled pair distribution,
\begin{equation}
Z  =  1 + \frac{1}{2N} \sum_{i \neq j}^{N} \delta(\frac{r_{ij}}{D_{ij}} - 1^{+})  \label{pressure}
\end{equation}
where $r_{ij}$ is the separation between particles $i$ and $j$, $D_{ij} = (\sigma_{i} + \sigma_{j}) / 2$, with $\sigma_i$ the diameter of particle $i$.
We perform simulations using different system sizes, $N=1024$, 4096 and 16384.

Our first task is to prepare equilibrium configurations of the system at large packing fraction, defined as $\phi = \pi\sum_i \sigma_i^2/(4L^2)$.
Following previous work~\cite{bcno2016,nbc2017}, we use a SWAP MC technique which combines translation MC moves for single particles and SWAP moves for randomly chosen pairs of particles.
To optimize the thermalization efficiency, the ratio of SWAP versus translational moves is set to 0.2~\cite{nbc2017}.

\begin{figure}
\includegraphics[width=0.48\textwidth]{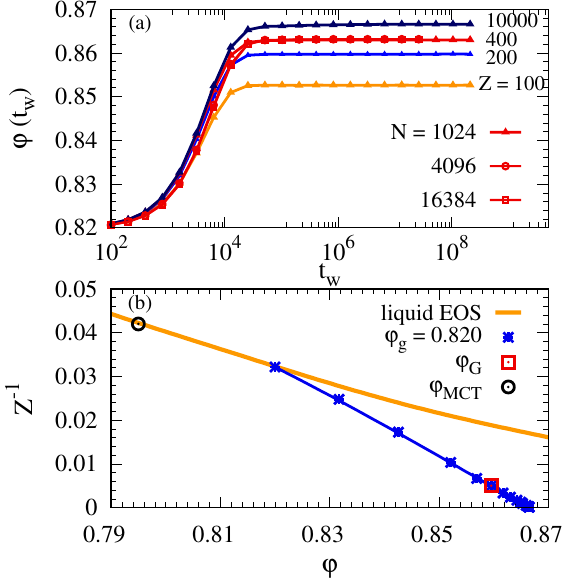}
\caption{\label{fig:fig1} (a) Waiting time evolution of the volume fraction $\phi$ after the pressure is suddenly increased from $Z_g=31$ to the targeted value for $N=1024$. For each pressure shown, a steady state value of the density is reached after $t_w \approx 10^4$. For $Z=400$ we show the compressions obtained for larger systems with $N=4096$ and 16384, which exhibit no system size dependence. (b) Equation of state for the equilibrium fluid, and glass equation of state obtained by compression of an initially equilibrated system at $\phi_g=0.820$. For reference, we show the location of the mode-coupling crossover, and of the avoided Gardner transition $\phi_G$.}
\end{figure}

At thermal equilibrium, the equation of state of the polydisperse hard disk fluid is well described by the following empirical equation of state~\cite{eos2001},
\begin{equation}
Z (\varphi) = \frac{1}{1 - \varphi} + \frac{M_{1}^{2}}{M_{2}} \frac{(1 + \varphi / 8)\varphi}{(1-\varphi)^{2}}
\label{eos},
\end{equation}
where $M_{1}$ and $M_{2}$ are the first and second moments of the diameter distribution. The measured equation of state in shown in Fig.~\ref{fig:fig1}.

By measuring the translational dynamics of the equilibrium supercooled fluid, we observe that the dynamics slows down rapidly as $\phi$ increases. By fitting the relaxation time to a power law divergence, we estimate the location of the mode-coupling crossover to be $\varphi_{MCT} = 0.795$, see Fig.~\ref{fig:fig1}. Above this density, particle diffusion becomes very slow when standard MC simulations are employed, but the SWAP MC method allows us to prepare much denser systems in equilibrium conditions. In this work, we prepare $N_s$ independent equilibrium configurations at $\phi_g=0.820$, where the pressure is $Z_g \approx 31$. At this packing fraction, the standard MC dynamics is nearly frozen and diffusive dynamics is not observed  over the duration of our simulations. Therefore, each of the $N_s$ configurations corresponds to a different glass sample.

Having prepared equilibrium glass configurations, we then employ $NPT$ standard MC simulations to rapidly compress the system.
In our simplest protocol, the pressure is instantaneously changed from its equilibrium value $Z_g=31$ to a much larger pressure $Z>Z_g$ to rapidly compress the system. This is analogous to an instantaneous temperature quench. We denote $t_w$ the `waiting time' spent since the pressure has been suddenly increased to its targeted value.

It is important to perform rapid compressions to avoid the formation of localised "defects"~\cite{sbz2017, 1d2018, defect2}, which correspond to small groups of particles which may undergo a small rearrangement in the initial steps of the compression when the pressure is not yet very large. Our simulations indicate that this happens with a low but finite probability, and this is distinct from the more collective Gardner physics we wish to analyse at larger pressures.

In order to obtain very fast compressions, we adjust the amplitude of individual compression moves in the $NPT$ MC protocol to obtain an acceptance rate of about 0.1. We also frequently attempt compression moves, once every 100 MC steps. The amplitude of the translational moves is carefully chosen to have an acceptance rate of about 0.6. Empirically, we find that these choices optimise both the measured compression rate and the CPU time. Examples of fast compressions from $Z_{g} = 31$ to $Z=100$, 200, 400 and 10000 are shown in Fig\ref{fig:fig1}. For each final pressure, we observe that after a short time of about $10^4$ MC sweeps, the density reaches a steady state value that increases with $Z$. The relation $Z(\phi)$ in that steady state defines the glass equation of state, which we report in Fig.~\ref{fig:fig1}.

As noted in a recent study~\cite{sz2018}, the efficiency of MC compression moves decreases rapidly as the system size increases. If one keeps the compression parameters fixed, this leads to an unwanted slowdown of the compression time, which scales proportionally to $N$. To keep the compression rate roughly constant, we introduce a slightly different compression move for larger systems in which we randomly select a subset of $N=1024$ particles (corresponding to the smallest system size studied in this work) at each compression move, and attempt to change their sizes. As can be seen in Fig.~\ref{fig:fig1}, this approach removes the $N$-dependence of the compression protocol. It is therefore much easier to compare results obtained for different system sizes.

In order to understand the structure of the landscape of hard disk glasses, we need to perform an average over the $N_s$ distinct glass samples produced in equilibrium, and for each glass sample, we also need to perform a thermal average over a large number $N_c$ of "clones". Each clone is prepared by replicating the particle positions of the glass at density $\phi_g=0.820$, but the different configurations are simulated using different random numbers for the MC moves, thus leading to independent trajectories within the same glass metastable basin. Hence, for each simulation, we need to choose a pair $(N_{s}, N_{c})$ for the number of glass samples and clones. To study macroscopic quantities, we have used two sets of values, $(N_s, N_c) = (100, 20)$ for general aspects of the physics related to the Gardner transition, while we use $(N_s, N_c) = (1, 100-1000)$ when we study individual glass samples.

\subsection{Physical observables}

In this section, we introduce the physical observables used throughout this work. Because we want to perform various levels of average, we need to be precise with the notations. We consider $N_s$ independent glass samples, that we design using Greek letters: $\alpha = 1, \cdots, N_s$. For each glass sample $\alpha$, we introduce $N_c$ clones, which we design using latin letters: $a=1,\cdots,N_c$. And we use $i=1,\cdots,N$ to design the particle index.

We then introduce two types of averages. We use an overline $\overline{O}$ to define a disorder average over the $N_s$ distinct glass samples, and use brackets $\langle O\rangle$ to denote averages over the clones within a given glass sample. To avoid complex notations, we use brackets for both averaging single clone quantities (over $N_c$ clones) and two-clone quantities (over $N_c(N_c-1)/2$ pairs of clones).

The basic observable we consider is the particle coordinate. We define
$\mathbf{R}^{\alpha}_{a, i} (t)$ the coordinate of particle $i$ within clone $a$ and sample $\alpha$ at time $t$. Tracking the motion at single particle level, we can measure the full mean-squared displacement (MSD),
\begin{equation}
D(t_{w}, t) =  \overline{\langle
\frac{1}{N} \sum_{i}|\mathbf{R}^{\alpha}_{a, i}(t_{w} + t) - \mathbf{R}^{\alpha}_{a, i}(t_{w})|^{2} \rangle} ,
\label{delta}
\end{equation}
where $t_{w}$ is the waiting time since the beginning of the compression.
It is important to keep the two-time notations, since the emergence of aging dynamics is a powerful tool to probe the Gardner transition~\cite{bcjpsz2016}.

There is an important complication which emerges in two-dimensional solids, which takes the form of long-range cooperative motion~\cite{2dsolid}. These so-called Mermin-Wagner fluctuations happen at finite temperature and pressure, and may dramatically affect the single particle dynamics we wish to study in the glass phase~\cite{2dglass,mw2017}. These fluctuations happen in addition to the vibrational motion within the cage, but they do not change the nature of the glass phase itself. To disentangle the true emergence of marginal glasses (which involve spatially correlated motion) from the more mundane Mermin-Wagner fluctuations, we introduce cage-relative coordinate as follows~\cite{mw2017,experiment2016}
\begin{equation}
\mathbf{r}^{\alpha}_{a, i}(t) = \mathbf{R}^{\alpha}_{a, i}(t) - \frac{1}{N_{i}} \sum_{j=1}^{N_{i}} \mathbf{R}^{\alpha}_{a, j}(t),  \label{cr_r}
\end{equation}
where $N_{i}$ is the number of nearest neighbors of particle $i$. We define neighbors as particles separated by a distance smaller than $2\overline{\sigma}$, which essentially includes the first shell of neighbors.
Using the cage relative coordinates, we can then define the cage-relative mean-squared displacement as
\begin{equation}
\Delta(t_{w}, t) = \overline{\langle
\frac{1}{N} \sum_{i} |\mathbf{r}^{\alpha}_{a, i}(t_{w} + t) - \mathbf{r}^{\alpha}_{a, i}(t_{w})|^{2} \rangle}.
\label{cr_delta}
\end{equation}
We also introduce the instantaneous MSD
$\Delta^{\alpha}_{a}(t_{w}, t)$ for a given clone $a$ within sample $\alpha$,
$\Delta^{\alpha}(t_{w}, t) = \langle \Delta^{\alpha}_{a}(t_{w}, t) \rangle$
for a given sample $\alpha$, such that
$\Delta(t_{w}, t) = \overline{\Delta^{\alpha}(t_{w}, t)}$.

\begin{figure}
\includegraphics[width=0.48\textwidth]{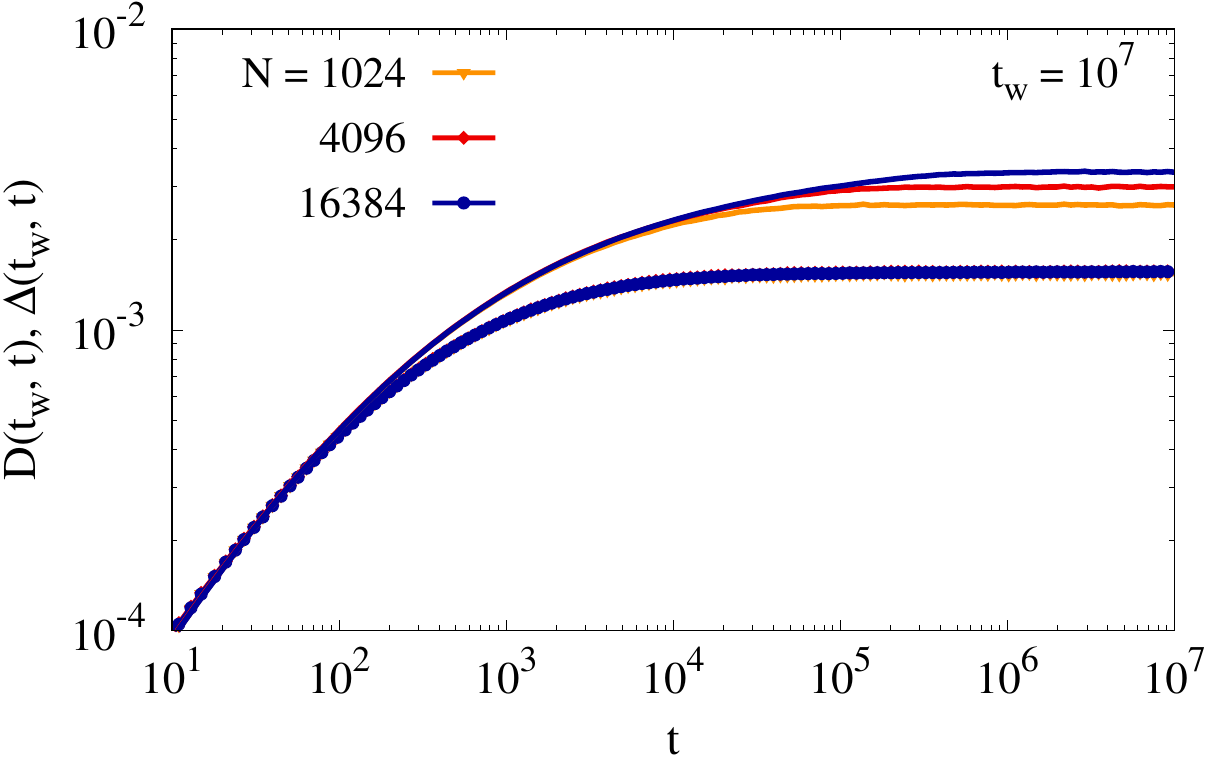}
\caption{\label{fig:fig2} Full MSD $D(t_{w}, t)$ (lines) and cage-relative MSD $\Delta(t_{w}, t)$ (symbols) measured at pressure $Z = 100$ for various system sizes $N = 1024$, $4096$, $16384$ with $t_{w} = 10^{7}$. The strong finite size effects due to Mermin-Wagner fluctuations are removed by using cage relative coordinates. We use cage-relative quantities for all measurements.}
\end{figure}

In Fig.~\ref{fig:fig2}, we show some representative data for both the full MSD $D(t_w,t)$ and the cage relative MSD $\Delta(t_w,t)$ after a quench from $Z_g=31$ to $Z=100$. Not only is the MSD much larger than the cage relative one, but its system size dependence is very different, as $D(t_w,t)$ seems to grow without bound as $N$ increases~\cite{mw2017}. By contrast, the cage relative MSD is insensitive to system size, showing that Mermin-Wagner fluctuations have properly been removed from our measurements.
In the rest of this work we quantify all the dynamic observables using cage relative definitions only. Whereas this distinction is crucial to quantify displacements (and, thus, dynamics), it is immaterial as far as particle positions are concerned.

The study of complex landscape makes heavy use of two-clone quantities, in the spirit of spin glass physics where the order parameter is an overlap between cloned configurations. In the present case, we introduce the distance between clones $a$ and $b$ as
\begin{equation}
\Delta_{AB} =
\overline{\langle
\frac{1}{N} \sum_i |\mathbf{r}^{\alpha}_{a, i}(t) - \mathbf{r}^{\alpha}_{b, i}(t)|^{2}
\rangle},
\end{equation}
Similarly to the MSD, we introduce the instantaneous value
$\Delta^{\alpha}_{ab}$ of the distance between clones $a$ and $b$ within the sample $\alpha$, and the typical distance between $\Delta^{\alpha}_{AB} =
\langle \Delta^{\alpha}_{ab} \rangle$ within a given sample $\alpha$.
From now on, we omit the time label for those one-time observables.

We are interested in the spatial fluctuations of the distance between clones. It is convenient to measure them in the Fourier domain and access the following structure factor,
\begin{equation}
S_{AB}(q) = \frac{1}{(\overline{\Delta_{AB}^\alpha})^2} \, \overline{
\langle \Delta^{\alpha}_{ab}(\mathbf{q}) \ \Delta^{\alpha}_{ba}(-\mathbf{q}) \rangle } ,
\label{sq1}
\end{equation}
where the distance field is defined in the Fourier domain as
\begin{equation}
\begin{split}
& \Delta^{\alpha}_{ab}(\mathbf{q}) =  \frac{1}{\sqrt{N}} \sum_{i} \Delta^{\alpha}_{ab, i} e^{ i{ \bf q} \cdot {\mathbf R}^{\alpha}_{a, i}}, \\
& \Delta^{\alpha}_{ba}(-\mathbf{q}) =  \frac{1}{\sqrt{N}} \sum_{i} \Delta^{\alpha}_{ab, i} e^{ -i {\bf q} \cdot {\mathbf R}^{\alpha}_{b, i}},
\label{sq2}
\end{split}
\end{equation}
where $\Delta^{\alpha}_{ab, i} = |\mathbf{r}^{\alpha}_{a, i}(t) - \mathbf{r}^{\alpha}_{b, i}(t)|^{2}$. Here both the indexes $a$ and $b$ range from $1$ to $N_c$, to the keep the symmetry. We have implicitly performed an angular average in Eq.~(\ref{sq1}). For visualisation purposes, we represent snapshots of the real space version of the displacement field,
\begin{equation}
\Delta_{ab}^\alpha({\bf R},t) = \sum_i \Delta_{ab,i}^\alpha(t)\delta ({\bf R} - {\bf R}_{a,i}^\alpha(t)), \\
\label{field}
\end{equation}

A simpler quantity measuring the global fluctuations of the distance field is to obtain the $q \to 0$ limit of the structure factor $S_{AB}(q)$, which corresponds to the following susceptibility~\cite{sbz2017}
\begin{equation}
\chi_{AB} = N \frac{ \overline{\langle (\Delta^{\alpha}_{ab})^{2} \rangle} - \overline{\langle {\Delta^{\alpha}_{ab}} \rangle}^{2} }{C_0}
\end{equation}
where the normalisation is given by
\begin{equation}
C_{0} = \frac{1}{N} \sum_{i}[\overline{\langle{\Delta^{\alpha}_{ab, i}}^{2}\rangle} - \overline{\langle\Delta^{\alpha}_{ab, i}\rangle}^{2}],
\end{equation}
to ensure $\chi_{AB} \sim \mathcal{O}(1)$ for a fully uncorrelated field.

It is also useful to decompose the susceptibility $\chi_{AB}$ into two distinct components, which, respectively, quantify the fluctuations within a sample
$\tilde{\chi}_{AB}$ and the sample-to-sample fluctuations
$\overline{\chi}_{AB}$, defined as follows:
\begin{equation}
\begin{split}
& \tilde{\chi}_{AB} = N \frac{\overline{\langle (\Delta^{\alpha}_{ab})^{2} \rangle} - \overline{{\langle {\Delta^{\alpha}_{ab}} \rangle}^{2}}} {C_{0}} , \\
& \overline{\chi}_{AB} = N \frac{\overline{{\langle {\Delta^{\alpha}_{ab}} \rangle}^{2}} - \overline{\langle {\Delta^{\alpha}_{ab}} \rangle}^{2}} {C_
{0}} . \\
\label{twochi}
\end{split}
\end{equation}
It is easy to verify that $\chi_{AB} = \tilde{\chi}_{AB} + \overline{\chi}_{AB}$.
Finally, the susceptibility within a single sample $\alpha$ is measured as,
\begin{equation}
\chi_{AB}^{\alpha} = \frac{\sum_{i,j} [\langle \Delta^{\alpha}_{ab, i}\Delta^{\alpha}_{ab, j} \rangle - \langle \Delta^{\alpha}_{ab, i} \rangle \langle \Delta^{\alpha}_{ab, j} \rangle] }{\sum_{i}[\langle{\Delta^{\alpha}_{ab, i}}^{2}\rangle - \langle\Delta^{\alpha}_{ab, i}\rangle^{2}]}
\end{equation}
As before, we can see that $\chi_{AB}^{\alpha} \sim \mathcal{O}(1)$ if the variables $\Delta^{\alpha}_{ab, i}$ for different particles $i$ are uncorrelated.

The final set of variables that we study in this work corresponds to performing a time average over the coordinates. To this end, we define time averaged cage relative positions for each particle after a quench:
\begin{equation}
\underline{\mathbf{r}}^{\alpha}_{a, i}(t_{w}, \tau) = \frac{1}{\tau} \int_{t_{w}}^{t_{w} + \tau} dt \ \mathbf{r}^{\alpha}_{a, i}(t) .
\label{ta_r}
\end{equation}
Here, we average the coordinates between time $t_{w}$ and $t_{w} + \tau$.
Computing the difference between clones, we have
\begin{equation}
\underline{\Delta}_{AB} (t_w,\tau)= \overline{\langle \frac{1}{N} \sum_{i}| \underline{\mathbf{r}}^{\alpha}_{a, i} - \underline{\mathbf{r}}^{\alpha}_{b, i}|^{2}
\rangle} .
\label{tadelta}
\end{equation}
The physical idea is that the long-time average will provide an estimate of the average position around which particle $i$ performs vibrations in clone $a$ within sample $\alpha$. We expect that all average positions would be the same from one clone to the next in a simple glass, whereas in a marginal phase characterized by a complex landscape, the average positions can be distinct for different clones. Finally, the time dependence of $\underline{\Delta}_{AB} (t_w,\tau)$ will inform us about how slow is the exploration of the landscape by the different clones within a given glass sample.

\section{Evidence of an (avoided) Gardner transition}

\label{evidence}

In this section, we use the system size $N=1024$, and perform averages over both samples and clones. We use $(N_{s}, N_{c}) = (100, 20)$ to obtain a reasonable statistical accuracy (requiring these numbers to be as large as possible) within the available computer resources~\cite{sg2011sample, sg2014, sz2018}. We follow earlier work and compress the system instantaneously using the protocol shown in Fig.~\ref{fig:fig1}, starting from the equilibrium state at $(\varphi_{g}, Z_{g}) = (0.820, 31)$ along the glass equation of state to pressures $Z = 100, \cdots, 10000$.

\subsection{Signatures of a Gardner crossover}

\begin{figure}
\includegraphics[width=0.48\textwidth]{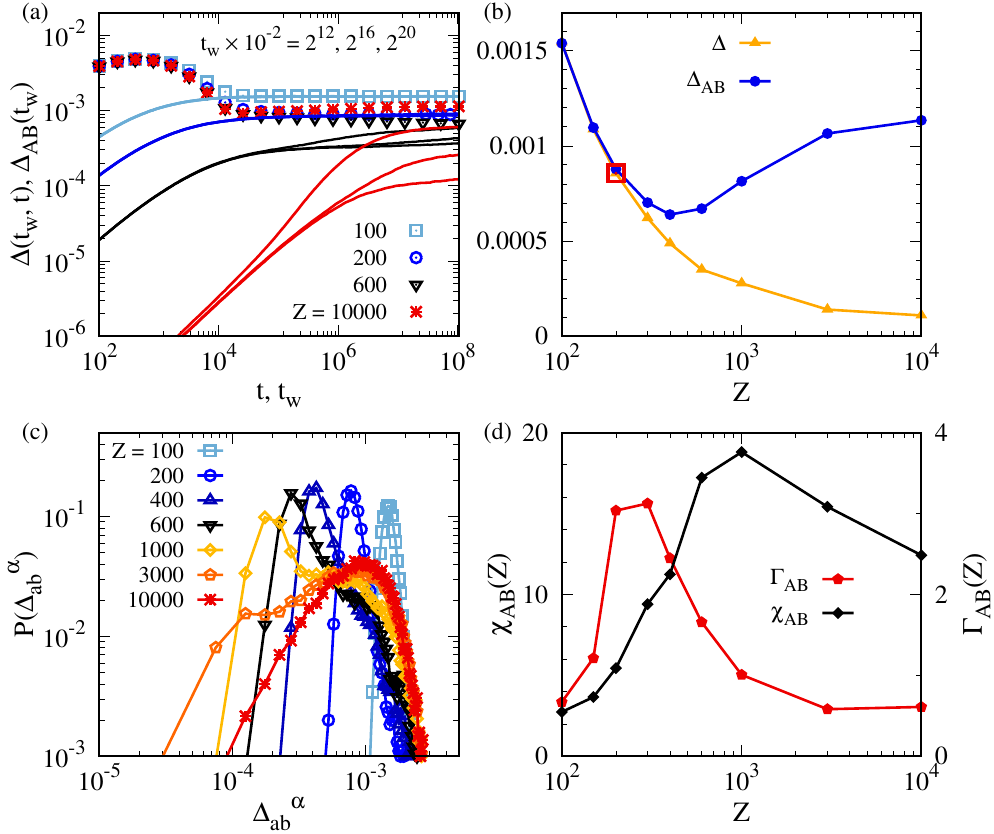}
\caption{\label{fig:fig3} Signatures of an (avoided) Gardner transition in $2d$ hard disks following the dynamics after a quench.
(a) Time evolution of $\Delta(t_{w}, t)$ (line) and $\Delta_{AB}(t_w)$ (symbols) for different pressures. For $Z > Z_G \approx 200$, $\Delta(t_{w}, t)$ exhibits aging and is smaller than $\Delta_{AB}(t_{w})$ at long times.
(b) The plateau values $\Delta(t_{w} \rightarrow \infty, t \rightarrow \infty)$ and $\Delta_{AB}(t_{w} \rightarrow \infty)$ start to differ for $Z > Z_{G}$.
(c) The probability distributions of distances between clones at various pressures for $t_w = 10^8$ become non-Gaussian for $Z>Z_G$.
(d) The skewness $\Gamma_{AB}(Z)$ and susceptibility $\chi_{AB}(Z)$ measured at $t_w = 10^8$ both exhibit a peak at a finite pressure.}
\end{figure}

In Fig.~\ref{fig:fig3} we reproduce several key signatures of the Gardner transition for the $2d$ hard disk glass, reported earlier for the $3d$ hard sphere system, which signal the emergence of marginal stability and a rough free energy landscape at large enough pressure~\cite{bcjpsz2016, sbz2017}.
For our $2d$ system, this can only signal a sharp Gardner crossover, not a real phase transition.

Our first observation is the evolution of the MSD $\Delta(t_w,t)$ and distance $\Delta_{AB}(t_w)$ following a quench, as shown in Fig.~\ref{fig:fig3} (a).
For small pressure, both quantities rapidly converge at long time to the same limit, indicating that the typical distance traveled by one clone is similar to the typical distance between clones. This shows that the structure of the glass remains simple. For pressures larger than $Z_G \approx 200$, by contrast, the MSD $\Delta(t_w,t)$ is strongly dependent on the waiting time, indicating slow dynamics. And the MSD no longer converges to the typical distance between clones, which indicates that the dynamics within the glass is no longer ergodic.

From these data, we compute the long time limit (in practice, we use $10^8$ MC sweeps) of both $\Delta$ and $\Delta_{AB}$ obtained after a quench. The results are shown in Fig.~\ref{fig:fig3} (b), as a function of the pressure where the quench is performed. As expected, we find that for $Z < Z_G$ both quantities converge to the same long-time limit, but depart gradually from one another as $Z$ increases beyond $Z_G$.

In Fig.~\ref{fig:fig3} (c), we show the probability distribution of distances
$P(\Delta^{\alpha}_{ab})$ measured at long waiting time $t_w=10^8$ and various pressures. The distribution has a simple Gaussian form for $Z < Z_G$, but becomes strongly non-Gaussian for $Z>Z_G$, which can be taken as another sign that a complex structure of the glass basin emerges beyond $Z_G$ characterized by a broad distribution of distances between the clones, as found before~\cite{bcjpsz2016}.
We observe that the distribution $P(\Delta^{\alpha}_{ab})$ first broadens as $Z \rightarrow 1000$, but then it narrows down again, as the pressure increases further, $Z > 1000$, which corresponds to a non-monotonic evolution of the susceptibility $\chi_{AB}(Z)$, as shown in Fig.~\ref{fig:fig3} (d). In addition, for intermediate pressures $Z = 200 \cdots 1000$, the distribution $P(\Delta^{\alpha}_{ab})$ develops a clear shoulder. This indicates that different pairs of clones can be either very close, or, instead, very distant from one another, and suggests already that some kind of self-organisation of the glass basin emerges near $Z_G$.

Finally, we also employ a procedure used before~\cite{bcjpsz2016} to locate the Gardner transition, and compute the skewness $\Gamma_{AB}(Z)$ of the distributions $P(\Delta_{ab}^\alpha)$, see Fig.~\ref{fig:fig3} (d). The skewness presents a clear peak, but we find that the position of the peak depends on the chosen timescale, shifting from $Z \approx 200$ at $t_{w} = 10^{7}$ to $Z \approx 300$ at $t_{w} = 10^{8}$ with no sign of a saturation.

To summarise, the results reported in Fig.~\ref{fig:fig3} indicate that all key signatures of the Gardner transition seem to survive in our $2d$ hard disk system, which appears to behave similarly to the $3d$ and mean-field systems~\cite{ckpuz2014, bcjpsz2016, sz2018}. Notice that some of these important signatures are absent for the nearly $1d$ system studied in Ref.~\cite{1d2018} and $3d$ and $2d$ systems of soft spheres~\cite{sbz2017, defect2}, where collective aging effects are not observed. At this stage, it is unclear that the Gardner transition is indeed avoided in our system, but the absence of a sharp phase transition will become more obvious in Sec.~\ref{hierarchical}.

\subsection{Growing time scales and length scales}

Next, we characterize the time scales and length scales associated to the emergence of a rough landscape inside hard disk glasses.

\begin{figure}
\includegraphics[width=0.48\textwidth]{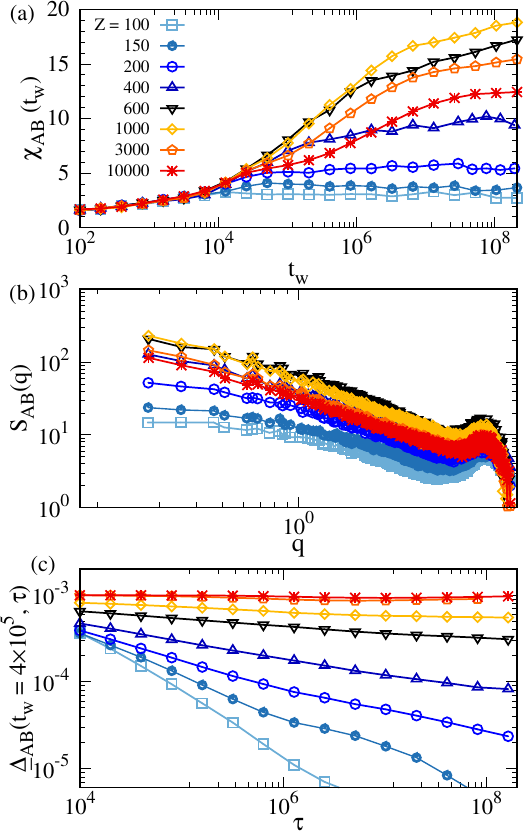}
\caption{\label{fig:fig4} Growing time scales and length scales in hard disk glasses.
(a) Time evolution of susceptibilities $\chi_{AB}(t_{w})$ for various pressures.
(b) $S_{AB}(q)$ at time $t_{w} = 10^{8}$ for various pressures indicated in (a).
As $q \rightarrow 0$, $S_{AB}(q)$ of $Z > Z_{G} \approx 200$ displays a growing peak, which suggests long-range fluctuations in real space over a length scale as large as $L/2 \approx 15$ for $Z > Z_G$.
(c) $\underline{\Delta}_{AB}(t_{w}, \tau)$ for pressures as in (a), with $t_{w} = 4\times10^{5}$. The increasingly slow decay at larger pressure directly reveals that the exploration of glass basin by individual clones becomes increasingly difficult at large pressure.}
\end{figure}

First, we follow the time evolution of the global fluctuations, $\chi_{AB}(t_w)$ in Fig.~\ref{fig:fig4} (a). For all pressures, the initial growth of $\chi_{AB}$ up to $t_w = 10^4$ is identical, which corresponds to the initial compression of the system. For $Z=100 \cdots 200$, the susceptibility rapidly reaches a constant limiting value, which increases by a factor of about 2. For $Z > 400$, no steady state value can be observed, and the susceptibility remains time dependent. It grows to reach large values of about 20 for $Z=1000$ with a slow time dependence at long times compatible with a nearly logarithmic increase with time (the data are shown in a log-lin representation in Fig.~\ref{fig:fig4} (a)).

Interestingly, for $Z=3000$ and 10000 the susceptibility becomes somewhat smaller, indicating a non-monotonic behaviour with $Z$, which was already apparent in Fig.~\ref{fig:fig3} (d). This overall behaviour is qualitatively similar to observations performed in $3d$ spin glass models~\cite{sz2018}.
Roughly speaking, the non-monotonicity corresponds to the competition between
the emergence of an increasingly complex landscape that tends to increase the susceptibility, and the slowing down of the dynamics which makes the exploration of that landscape more difficult at large pressures.

A similar non-monotonic behaviour can be observed in real space, as shown in Fig.~\ref{fig:fig4} (b) which represents the spatial fluctuations of the distance field $S_{AB}(q)$ for a time $t_w=10^8$ and various pressures.
For each pressure, the structure factor has a peak near $q \approx 6$, which corresponds to the interparticle distance. More interesting is the development of a peak at $q \rightarrow 0$, which corresponds in real space to spatial fluctuations happening at length scales much larger than the interparticle scale. As $Z$ increases, this peak becomes higher, and a saturation to a low $q$ plateau, which is clearly visible for $Z<200$ becomes difficult to observe. This directly indicates that the fluctuations of the distance field between clones inside glass basins become spatially correlated over a large correlation length that grows as the pressure increases. Just as for the susceptibility, we observe that the height of a low-$q$ peak has a non-monotonic behaviour with $Z$, reaching its maximum near $Z=1000$. For $N=1024$, the linear size of the system is $L\approx 30$, and so the observations in Fig.~\ref{fig:fig4} (b) indicate that a collective correlation length of at least $L/2 \approx 15$ develops in the system near $Z_G$. Given the relatively modest range of wavevector at hand, it is difficult to quantitatively extract the value of a correlation length from these data, a task which is known to be difficult in glass-formers~\cite{berthier2011dynamical}. We will nevertheless confirm these observations using larger systems in Sec.~\ref{larger}.

We now turn to the growing time scales associated to the signatures of a Gardner transition. In Fig.~\ref{fig:fig4} (c) we report the evolution of the distance $\underline{\Delta}_{AB} (t_w,\tau)$ between time-averaged positions, see Eq.~(\ref{tadelta}). We choose $t_w = 4\times10^5$, which is large enough for the density to converge to its steady state value. For $Z =100 \cdots 200$, the distance decays to zero at long times, indicating that the structure of the glass basin remains simple at those pressure, and all clones can freely explore the entire basin. However, the time scale for this exploration increases rapidly with $Z$, from $\tau \sim 10^6$ at $Z=100$ to $\tau > 10^8$ at $Z = 200$. For even larger pressures, $\underline{\Delta}_{AB} (t_w,\tau)$ does not decay to zero, indicating that different clones are now confined in different restricted part of the glass basin that are not dynamically accessible at these pressures. In Sec.~\ref{single} we revisit the physical meaning of this increasing time scale in terms of emerging barriers that appear near $Z_G$.

\subsection{Thermal average versus disorder average}

\begin{figure}
\includegraphics[width=0.48\textwidth]{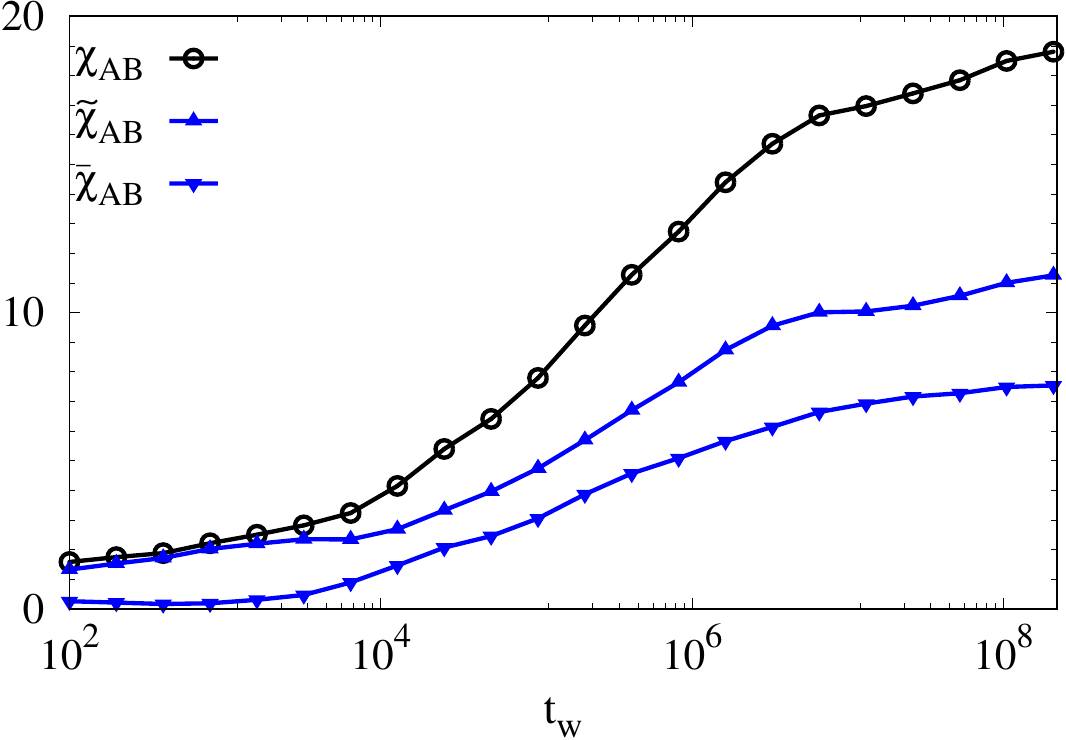}
\caption{\label{fig:fig5}
Time evolution of the susceptibilities at $Z = 1000$. The total susceptibility $\chi_{AB}$ is decomposed into sample-to-sample fluctuations $\overline{\chi}_{AB}(t_{w})$ and clone-to-clone fluctuations within each individual sample $\tilde{\chi}_{AB}(t_{w})$. Both contributions are comparable in amplitude and time evolution.}
\end{figure}

In the quantities presented above, the fluctuations have two distinct origins: there are clone-to-clone fluctuations within each glass sample, and there are sample-to-sample fluctuations because each glass corresponds to a distinct particle packing.

We have performed a more detailed study of a small number ($N_s =10$) of individual glass samples, using for each glass a large number of clones, $N_c =100$. We find that each glass behaves dynamically similarly, and variations between samples are mostly quantitative. In particular, each individual sample exhibits a clear aging behaviour, indicating that for all samples, a non-trivial collective dynamics emerges at large enough pressures. Strikingly, all the key signatures of a Gardner transition shown in Fig.~\ref{fig:fig3} can be reproduced for a single glass.

However, the precise location of the crossover pressure differs from one sample to another. For $N=1024$, the range of these variations is estimated to vary from $Z_G \approx 100$ to $Z_G \approx 600$, for a set of glass configurations drawn from our pool of equilibrium systems at $(\varphi_{g}, Z_{g}) = (0.820, 31)$.

In order to estimate the relative influence of the two sources of the fluctuations, we decompose the susceptibility $\chi_{AB}$ into its two contributions defined in Eq.~(\ref{twochi}). For a quench to $Z=1000$, we show the result in Fig.~\ref{fig:fig5}. The data in Fig.~\ref{fig:fig5} indicate that the two sources of fluctuations are of the same order, and we find in addition that the ratio $\overline{\chi}_{AB} / \tilde{\chi}_{AB}$ increases only weakly with the pressure.

This result suggests that the analysis of single glass samples provides an important part of the fluctuations associated to the avoided Gardner transition. In addition, averaging over distinct glass basins washes out the detailed information related to the structure of the landscape within each glass.
In the next two sections, we thus perform a more detailed analysis of the evolution of the landscape in a single glass basin.


\section{Detailed analysis of a single glass sample}

\label{single}

In this section, we analyse the physics associated to the Gardner transition for a single glass sample selected randomly. The protocol and parameters are the same as in Sec.~\ref{evidence}. We generate $N_c= 1000$ clones to understand better the structure of this particular glass basin. For this sample, the key signatures shown in Fig.~\ref{fig:fig3} are reproduced with $Z_{G} \approx 200$. We first analyse the physics in the vicinity of the Gardner crossover near $Z_G$ and then move to larger pressures.

\subsection{Emergence of sub-basins at Gardner crossover}

\label{emergence}

\begin{figure}
\includegraphics[width=0.48\textwidth]{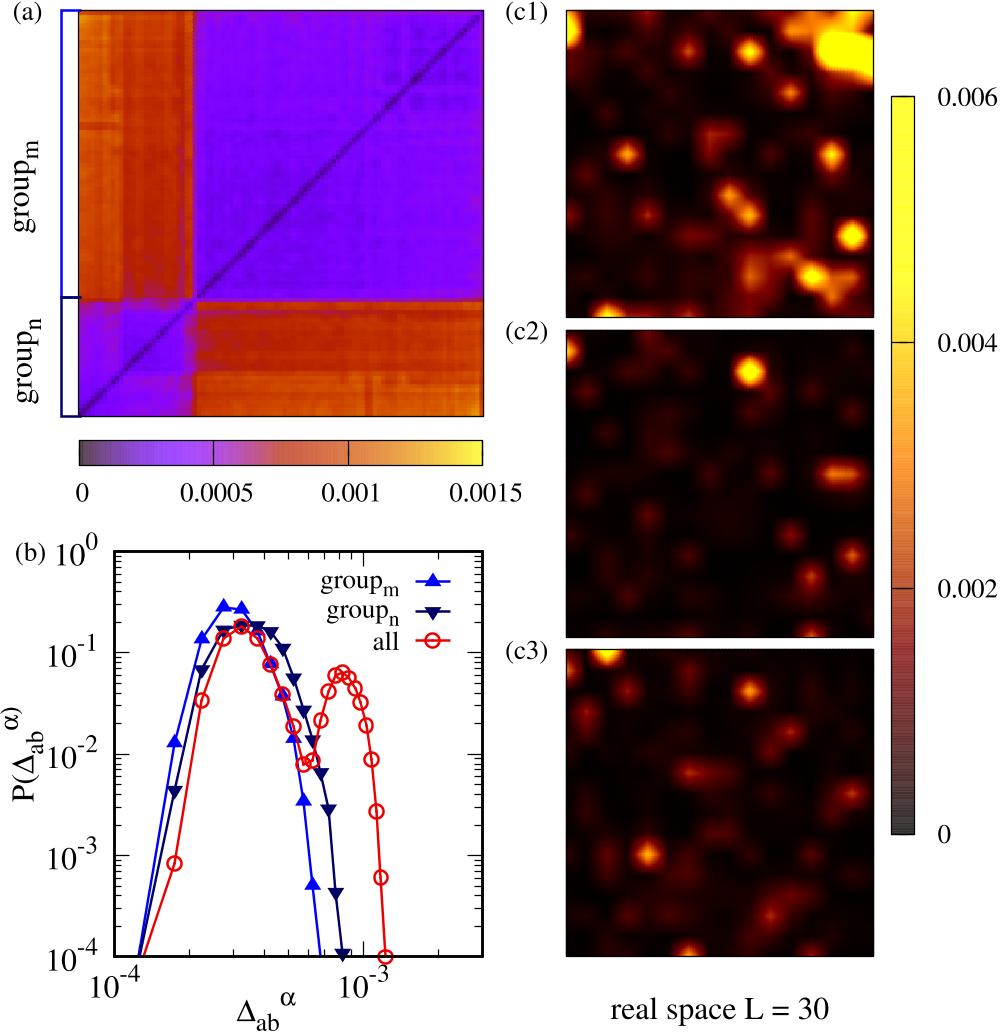}
\caption{\label{fig:fig6} Emerging complexity in the landscape of individual sample $\alpha$ with $N = 1024$ for $Z = 600 > Z_G$.
(a) Heat map of cage-relative MSD at $t_{w} = 2^{18}\times100$, with the clone index ranging from $1$ to $100$ after clustering. The clones can be divided into two subpopulations, group$_m$ and group$_n$.
(b) Corresponding probability distributions $P(\Delta_{ab} ^{\alpha})$ for all clones, group$_m$ and group$_n$.
(c1)-(c3) Cage-relative displacement field for clones belonging to different groups (c1), group$_m$ (c2), and group$_n$ (c3).}
\end{figure}

To understand the emerging structure of the glass basin, we use a large number $N_c$ of clones issued from the same equilibrium configurations at $Z_g=31$ and we independently compress them instantaneously at some pressures $Z>Z_G$.
After a given waiting time, we then measure the relative distance $\Delta_{ab}^\alpha$ between clones $a$ and $b$ for this sample $\alpha$. It is useful to cluster the clones according to their relative distances. With the simple procedure explained in Appendix~\ref{app:map}, we can relabel the clones and construct heat maps such as the ones shown in Fig.~\ref{fig:fig6} (a), where the color is coding for the distance between the clones. For sample $\alpha$ in the condition of $Z = 600$ and $t_{w} = 2^{18}\times100$, the map reveals a clustering of the clones into two subpopulations, which we refer to as group$_m$ and group$_n$. We present maps for different times and pressures below. This indicates that, after a fast compression, the different clones now explore two distinct parts of the glass basin that are not easily accessed dynamically. The underlying physical picture is that of a glass basin that is now fractured into two sub-basins separated by a barrier between them.

We now measure the corresponding probability distribution functions $P(\Delta_{ab}^\alpha)$ shown in Fig.~\ref{fig:fig6} (b). As expected from the structure of the heat map in Fig.~\ref{fig:fig6} (a), the full distribution of distances between clones is bimodal, which reflects the clustering of the clones into two families. Clones within a given group are close to each other, while they are relatively far from all clones belonging to the other group, which leads to two peaks in the distributions. Decomposing the clones into group$_m$ and group$_n$, we also measure the distribution of distances between clones belonging to the same group in Fig.~\ref{fig:fig6} (b). These distributions are nearly Gaussian, centered around a small value roughly equal to the cage size at the pressure $Z = 600$.

Finally, in Fig.~\ref{fig:fig6}(c1-c3), we visualise the displacement fields between clones, $\Delta_{ab}^\alpha({\bf R})$, generated by choosing a pair of clones belonging to group$_m$ and group$_n$ as in Fig.6 (c1), or to the same group$_m$ (Fig.6 (c2)) or group$_n$ (Fig.6 (c3)).
The method to construct these snapshots is detailed in Appendix ~\ref{app:snapshot}. We observe long-range correlations in Fig.6 (c1) which are instead much shorter-ranged in panels Fig.6 (c2-c3). These snapshots suggest that the emergence of large length scale across the Gardner crossover $Z_G$ is a direct consequence of the fracturation of the glass basin into multiple sub-basins.

\subsection{Physical interpretation of aging dynamics}

We now extend the analysis of the heat map to a larger range of timescales and pressures. Our goal is to explore the connection between the emergence of barriers within the glass basin and the aging dynamics reported in Fig.~\ref{fig:fig3} (a) and the growing time scale reported in Fig.~\ref{fig:fig4} (c).

\begin{figure}
\centering
\includegraphics[width=0.48\textwidth]{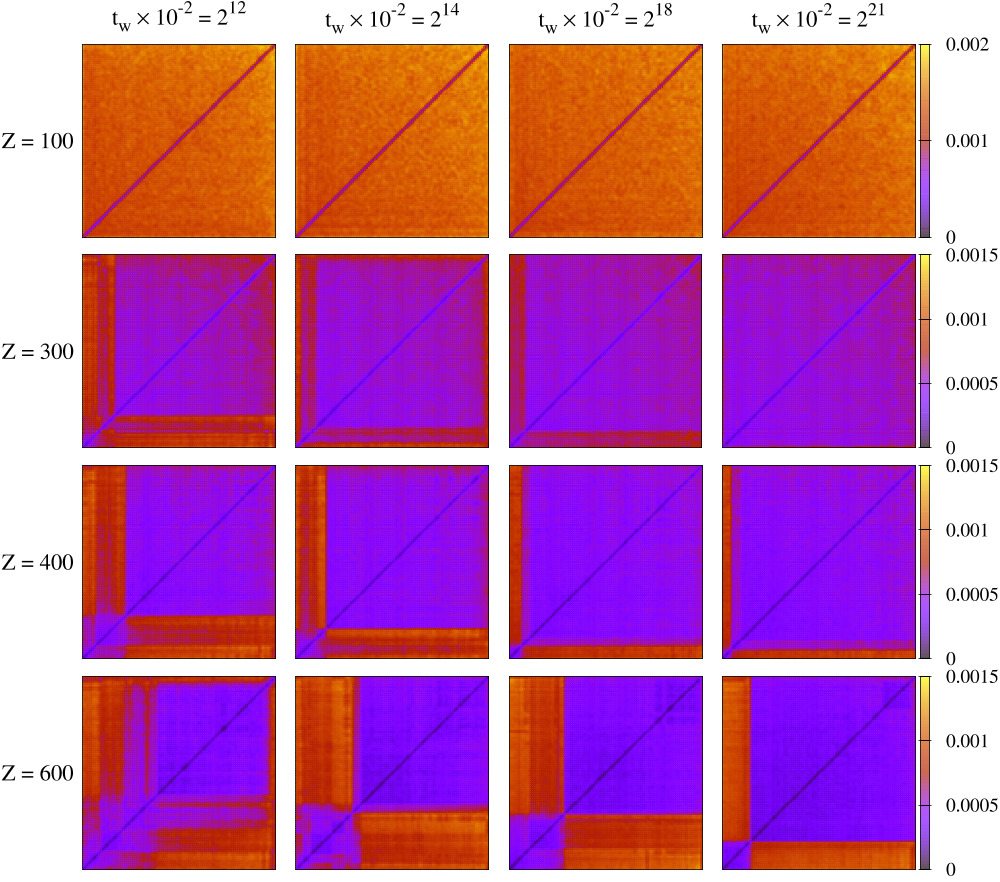}
\caption{\label{fig:fig7} Heat maps of distances between clones $\Delta_{ab} ^{\alpha} (t_{w})$ for sample $\alpha$ with $N=1024$, organized by $N_c = 100$ clones. The waiting time increases from left to right, the pressure from top to bottom, as indicated. Whereas the maps at $Z = 100 < Z_{G}$ are featureless and time-independent, they exhibit an organisation into two sub-basins for $Z \geq 300$, which slowly disappears as $t_w$ increases.}
\end{figure}

We first display the evolution of the heat maps for several pressures $Z$ and various waiting times $t_w$ in Fig.~\ref{fig:fig7}. The first row of heat maps shows the results obtained for a relatively low pressure, $Z=100 < Z_G$, where aging is absent and the vibrational dynamics is featureless.
Accordingly, we observe that all clones within the glass sample resemble each other and can freely explore the basin. As a result, the heat map is both featureless and time independent. For $Z=300$, 400 and 600, one can see the emergence of the two subpopulations group$_m$ and group$_n$ discussed in Sec.~\ref{emergence}. Fixing the pressure, the time dependence observed shows that as $t_w$ increases the size of the group$_m$ enlarges, whereas group$_n$ shrinks. In addition, this dynamics slows down as $Z$ increases so that for a fixed $t_w$, the number of clones surviving within group$_n$ increases with $Z$.

\begin{figure}
\includegraphics[width=0.48\textwidth]{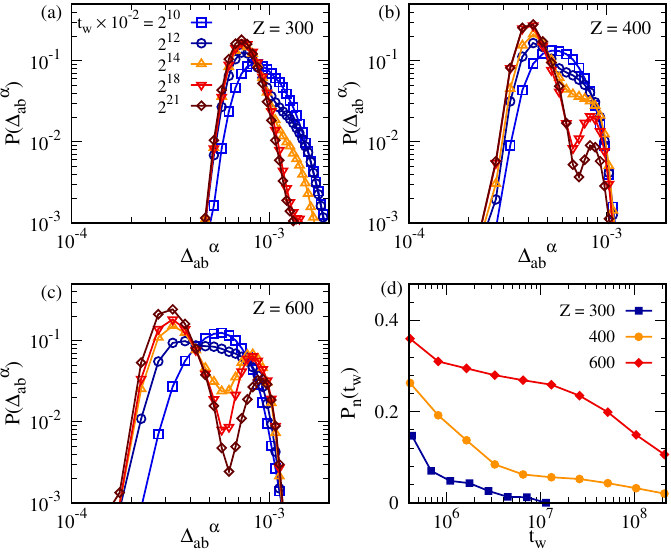}
\caption{\label{fig:fig8} (a)-(c) Time evolution of the probability distribution $P(\Delta_{ab} ^{\alpha})$ for a single sample $\alpha$ with $N = 1024$ and various pressures, measured with $N_c=1000$ clones.
(d) The time evolution of the fraction of clones belonging to the less stable group$_n$ for different pressures resembles the dynamic data in Fig.~\ref{fig:fig4} (c).}
\end{figure}


The emergence of a bimodal structure that slowly fades away with time is confirmed in Fig.~\ref{fig:fig8} (a-c), where we report the probability distributions function corresponding to the maps shown in Fig.~\ref{fig:fig7}.
In agreement with the maps, the distributions develop a bimodal structure at intermediate waiting times, where the peak at small distance corresponds to the amplitude of thermal vibrations, whereas the peak (or shoulder for $Z=300$) at large distance corresponds to the typical distance between the two sub-basins identified by group$_m$ and group$_n$. For a fixed $t_w$, increasing the pressure tends to broaden the distributions, which is related to the growth of the susceptibility $\chi_{AB}$ near $Z_G$ shown in Fig.~\ref{fig:fig3} (d).

The time evolution of these bimodal distributions can be easily described by computing the evolution of the fraction of clones $P_n(t_w)$ that survive in the less stable group$_n$ after a time $t_w$. This can be done when the pressure is in the vicinity of the Gardner crossover.
We report the time evolution of $P_n(t)$ in Fig.~\ref{fig:fig8} (d).
In agreement with the time evolution of the maps and of the distributions, we find that group$_n$ becomes empty after $t_w \approx 10^7$ for $Z=300$, while the corresponding population survives much longer and remains finite even after $t_w = 2\times10^8$ for $Z=600$, suggesting that the dynamic evolution from one basin to the other becomes slower as the pressure increases.
The time decay of $P_n(t_w)$ is quantitatively similar to the evolution of the distance between average positions $\underline{\Delta}_{AB}(t_w,\tau)$ discussed in Fig.~\ref{fig:fig4} (c).

\begin{figure}
\includegraphics[width=0.48\textwidth]{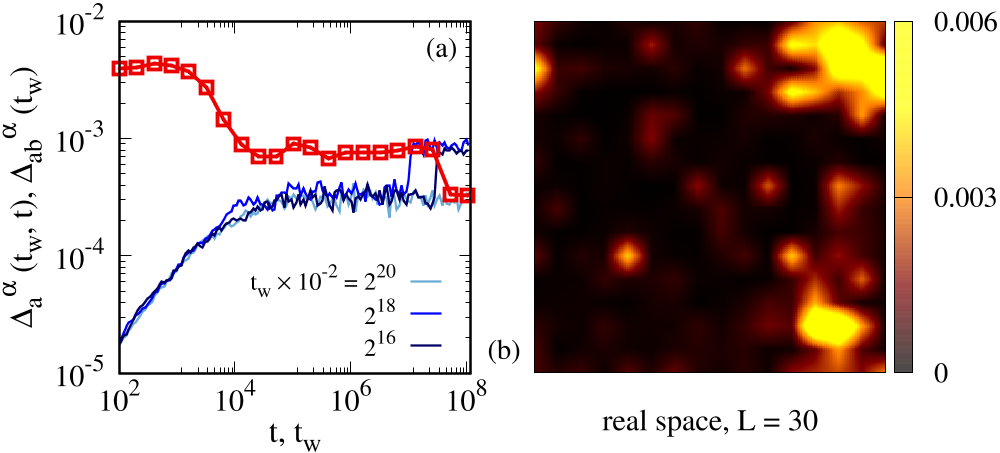}
\caption{\label{fig:fig9} Connection between landscape structure and aging dynamics for sample $\alpha$ at $Z=600$.
(a) Blue lines: aging dynamics for a single clone $a$ undergoing activated event from group$_n$ to group$_m$ via MSD $\Delta^{\alpha}_a(t_w,t)$.
Red square: the distance $\Delta^{\alpha}_{ab}(t_w)$ between clone $a$ and a randomly chosen clone belonging to group$_m$ jumps discontinuously after the rearrangement.
(b) The cage-relative displacements by the configurations of clone $a$ measured before and after the activated event detected in (a). This displacement field is very similar to the snapshot of $\Delta_{ab}^\alpha$ in Fig.~\ref{fig:fig6} (c1).}
\end{figure}

We now close the loop between the structure of the glass landscape, the emerging barriers near $Z_G$, and the physical interpretation of the aging dynamics arising for $Z > Z_G$.
A careful analysis of the MSD at the single clone level reveals that the existence of two subpopulations, group$_m$ and group$_n$, implies also two types of MSD.  Clones that belong at time $t_w$ to the more stable group$_m$
are characterized by simple MSD smoothly increasing to reach the cage size, and never leave that long-time plateau. These clones do not contribute to the aging dynamics.
By contrast, clones belonging at time $t_w$ to the less stable group$_n$ may undergo some activated event at time $t+t_w$ later to join group$_m$.
The example of such a clone, clone $a$, is shown in Fig.~\ref{fig:fig9} (a) (blue lines). The MSD first grows to the plateau corresponding to the cage size, but then a rearrangement takes place that increases the MSD further to a second plateau value. Notably, this second plateau corresponds to the typical distance $\Delta_{ab}^\alpha$ between group$_m$ and group$_n$ (red squares). After the rearrangement, the clone $a$ belongs to group$_m$.

Obviously, the time when the sudden rearrangement takes place in clone $a$ fluctuates from one clone to another. Therefore, after averaging over clones and over glass samples, the MSD eventually exhibits the smooth aging behaviour reported in Fig.~\ref{fig:fig3} (a).

We also confirm that the spatial correlations that develop during the aging are also directly controlled by the underlying structure of the phase space. We show in Fig.~\ref{fig:fig9} (b) a snapshot representing the real space displacement field corresponding to the activated event detected in clone $a$ in Fig.~\ref{fig:fig9} (a), by measuring $\Delta^{\alpha}_a(t_w,t)$ for well-chosen times. This displacement field is very similar to the distance field shown in the snapshot of Fig.~\ref{fig:fig6} (c1), corresponding to $\Delta_{ab}^\alpha$ between clones belonging to group$_m$ and group$_n$.
The remarkable similarity between Fig.~\ref{fig:fig9} (b) and Fig.~\ref{fig:fig6} (c1) indicates that the long-time aging dynamics in the vicinity of the Gardner crossover is fully controlled by activated relaxations between the emerging sub-basins.

Remarkably, this aging dynamics is spatially correlated over a relatively large length scale, comparable to the system size $L/2 \approx 15$, and involves a large number of particles. Thus, we conclude that the sub-basins structure that we detect for $2d$ hard disks, is qualitatively distinct from the highly localised defects found in both nearly $1d$ hard disks and $3d$ and $2d$ soft spheres~\cite{sbz2017, 1d2018, defect2}.

In summary, the detailed analysis of a single glass sample shows that as $Z$ increases across $Z_G$, the glass basin fractures into two sub-basins separated by a  collective barrier. After a sudden quench, clones are initially randomly trapped into one of the sub-basins, and as $t_w$ increases, the clones belonging to the less stable sub-basin undergo a collective, spatially correlated rearrangement to join the more stable sub-basin. This physical picture explains the emergence of growing timescales and lengthscales in the vicinity of the Gardner crossover. In particular, if fully accounts for the growing timescale detected directly in Fig.~\ref{fig:fig4} (c).

\subsection{Proliferation of states at large pressures}

\begin{figure}
\includegraphics[width=0.48\textwidth]{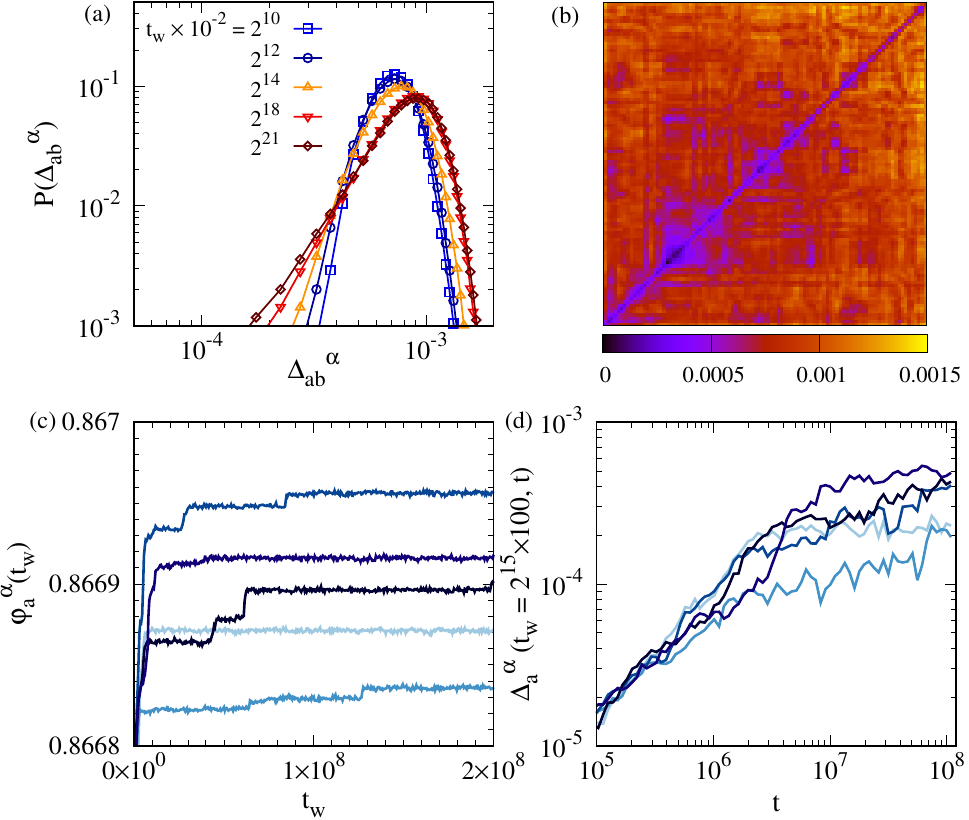}
\caption{\label{fig:fig10} Exploration of the landscape of sample $\alpha$ with $N=1024$ compressed at $Z=10000$.
(a) Time evolution of probability distribution $P(\Delta_{ab}^{\alpha})$.
(b) Corresponding heat map at $t_{w} = 2^{21}\times100$.
(c) Diversity of activated events in individual clones are captured by small jumps in the packing fraction.
(d) Corresponding evolution of the individual MSD at $t_w = 2^{15}\times100$.}
\end{figure}

We now present results of the same sample $\alpha$, obtained by quenching to much larger pressure, $Z = 10000 \gg Z_{G} \approx 200$. We again use $N_c = 1000$ clones and record the time evolution of the probability distribution function $P(\Delta_{ab}^\alpha)$; see Fig.~\ref{fig:fig10} (a). The distribution shows a peak at a large value (much larger than the cage size), and it starts to develop a tail at smaller values as $t_w$ increases. However, the simple bimodal structure found at smaller pressures is now totally absent. Physically, this means that after a quench to a large pressure, all clones are trapped within a large number of distinct structures, so that two clones chosen at random are typically separated by a large distance.
Thus, instead of finding two sub-basins separated by a single large barrier, the $N_c$ clones in sample $\alpha$ are now trapped in a number of ${\cal O}(N_c)$ distinct states. Correspondingly, when we perform a cluster analysis of the clones to construct a heat map, we find no interesting structure emerging from the map, see Fig.~\ref{fig:fig10} (b). We conclude that the number of basins increases dramatically upon approaching jamming.
The crossover between the simple bimodal structure observed for $Z=600$ in Fig.~\ref{fig:fig8} and the very complex structure for $Z=10000$ in Fig.~\ref{fig:fig10} is not sudden: the change of pressure between these two state points is quite large (a factor of 16), and quenches performed at pressure values intermediate between those reveal a gradual change from one limit to the other.

Tracking the time evolution of the clones, we observe that there is still some activated dynamics taking place, which is responsible for the aging dynamics observed at the level of the averaged MSD. If one follows the time evolution of the packing fraction in individual clones, as shown in Fig.~\ref{fig:fig10} (c), one can observe that the density performs small jumps corresponding to activated events taking place within each clone. Similarly, the time evolution of the MSD for individual clones shown in Fig.~\ref{fig:fig10} (d) shows that the MSD exhibit small jumps, which presumably correspond to crossing small barriers between neighboring states. The amplitude of these jumps is smaller than at lower pressure, suggesting that aging at large pressures corresponds to crossing smaller barriers.

In summary, a direct quench to large pressures reveals a proliferation of accessible states, indicating that the glass basin breaks at large $Z$ into many sub-basins. Dynamically, the clones at such large pressure can only cross small barriers and explore a very small part of the entire glass basin. Altogether, these results therefore point towards a hierarchical organisation of the glass landscape where the large sub-basins that appeared at smaller pressure are themselves broken into many smaller sub-basins, involving therefore a hierarchy of barriers.

The analysis of this complex structure emerging in a single glass sample is not easy. To get more insight into this structure, we turn to more complicated compression protocols to more evidently reveal the hierarchical structure of the landscape of hard disk glasses.


\section{Evidence of hierarchical landscape approaching jamming}

\label{hierarchical}

In the previous section, we observed a fracturation of the glass basin of sample $\alpha$ in two basins near the Gardner crossover, and a proliferation of dynamically inaccessible states at larger pressure. In this section, we demonstrate that the glass basin is actually organised in a hierarchical way where, as the pressure increases, the main basin breaks into sub-basins that themselves break into several sub-basins, etc. This description suggests the existence of a broad distribution of barriers separating these different states.

\subsection{Compressing the system in several steps}

\begin{figure}
\includegraphics[width=0.48\textwidth]{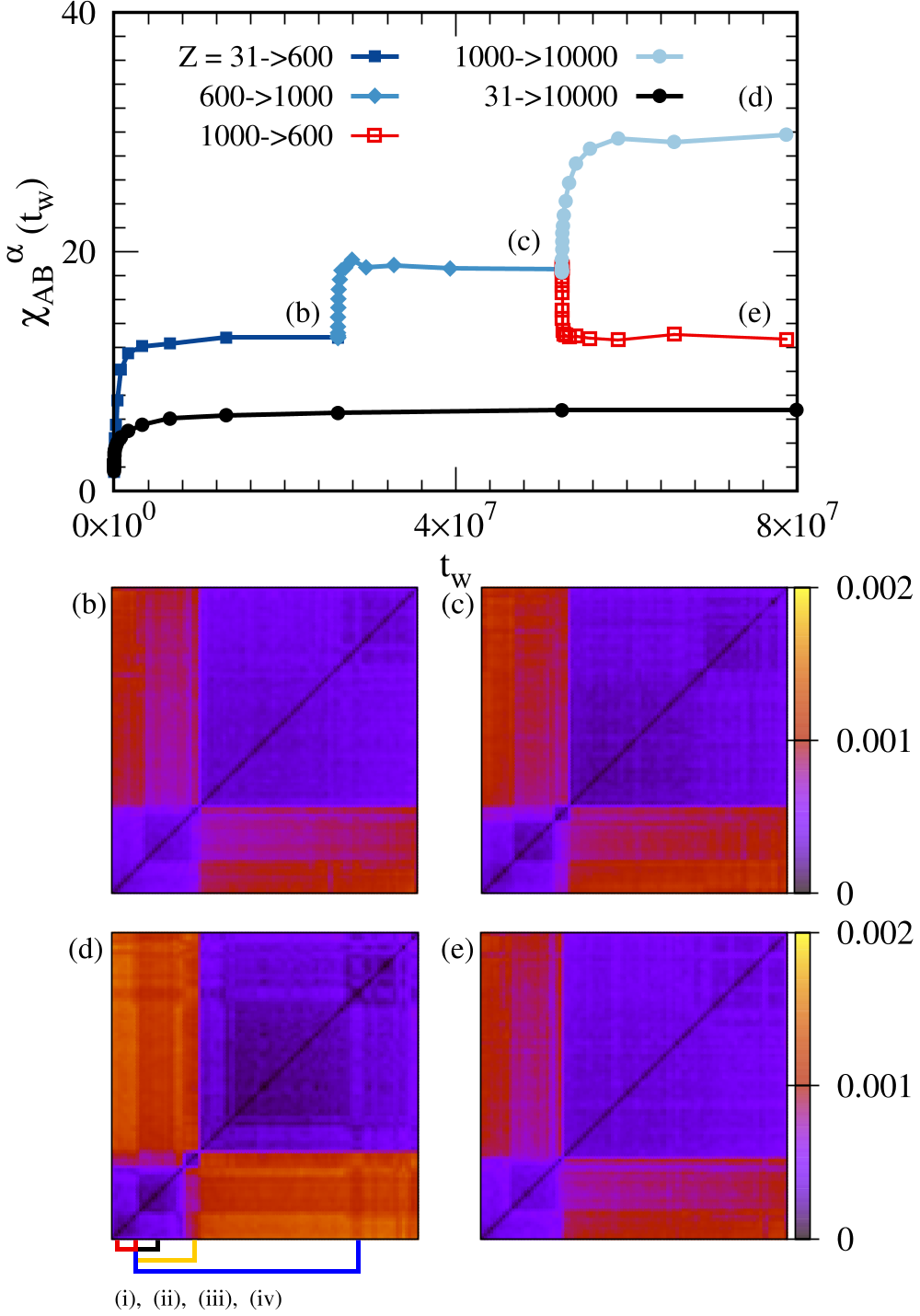}
\caption{\label{fig:fig13} Hierarchical landscape of sample $\alpha$ revealed by a multi-step compression protocol.
(a) Time evolution of the susceptibility $\chi_{ab}^{\alpha}(t_{w})$, in the various histories indicated by keys.
(b)-(e) Corresponding heat maps of distances $\Delta_{ab} ^{\alpha}$ for $N_c=100$ clones, at times indicated in panel (a). In particular, in (d) the map at $Z=10000$ clearly reveals a hierarchical structure with states within states. The selected pairs of clones (i-iv) in (d) are used to construct the snapshots in Fig.~\ref{fig:fig14}.}
\end{figure}

To illustrate the finer structure of the glass landscape, we continue our exploration of a single representative sample $\alpha$, but change our protocol. As an example, we use the following history: $Z = 31 \mapsto 600 \mapsto 1000 \mapsto 10000$, where the applied pressure is kept constant for a duration $dt = 2^{18}\times100$ at each step. In Fig.~\ref{fig:fig13} (a) we present the evolution of the susceptibility $\chi_{ab}^\alpha$ measured in sample $\alpha$ during this multi-step protocol, and the corresponding heat maps of distance in Fig.~\ref{fig:fig13} (b-e).

During the first stage $Z = 31 \mapsto 600$, we recover the physics obtained during a direct quench across $Z_G$, namely a slow growth of the susceptibility associated to a bimodal organisation of the states, as discussed in Sec.~\ref{single}.
If we waited a larger time at this pressure, group$_n$ would eventually disappear and all clones would be located inside group$_m$, which is the most stable sub-basin. Such a protocol is used below in Sec.~\ref{avoided}. Here instead, we again increase the pressure to $Z=1000$ after a waiting time $dt$ when both group$_m$ and group$_n$ are still populated, followed by another quench to $Z=10000$ at time $t_w = 2dt$. In this multi-step protocol, we find that the susceptibility increases rapidly after each of the steps, indicating that the distribution of distances between clones broadens at each step.

The most striking observation comes from the corresponding heat maps in Fig.~\ref{fig:fig13} (c-e), which show that the bimodal structure observed at $Z=600$ fractures into a finer structure hierarchically. Specifically, we observe that each blue square from Fig.~\ref{fig:fig13} (b) at $Z=600$ becomes decomposed into sub-squares at $Z=1000$ in Fig.~\ref{fig:fig13} (c), which themselves break into smaller clusters at $Z=10000$ in Fig.~\ref{fig:fig13} (d).

\begin{figure}
\includegraphics[width=0.48\textwidth]{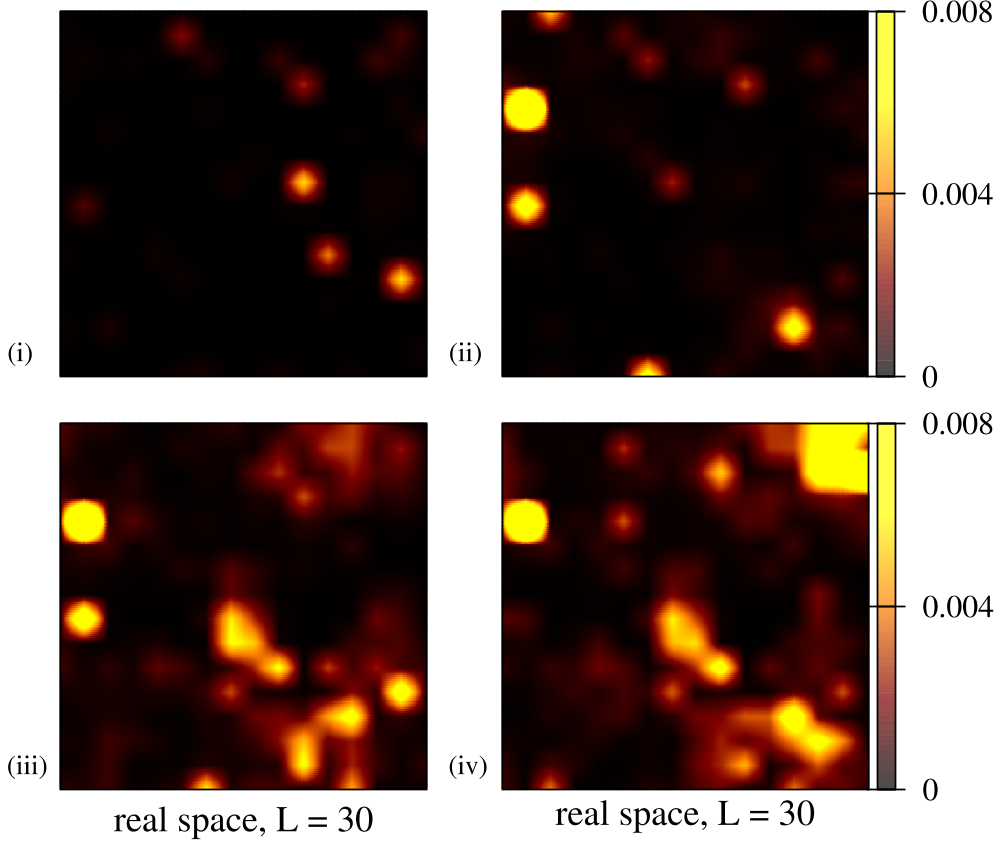}
\caption{\label{fig:fig14} Distance fields $\Delta_{ab} ^{\alpha} (\mathbf{R})$ for sample $\alpha$ at $Z = 10000$, using the four pairs of clones shown in Fig.~\ref{fig:fig13} (d). From (i) to (iv), the various levels of the hierarchy of states are associated to larger displacements and longer-ranged spatial correlations.}
\end{figure}

Whereas the clones were scattered across many different states with little organisation between them after a direct quench to $Z=10000$ (recall the heat map in Fig.~\ref{fig:fig10} (b)), the states reached at the same pressure in a multi-step compression protocol are now clearly clustered and organised in a hierarchical manner in Fig.~\ref{fig:fig13} (d). Physically the difference is that the subpopulation of clones in different sub-regions of the landscape are better thermalised in a multi-step protocol than by a direct quench to a large pressure, where the clones are instead trapped in a myriad of states and can barely evolve dynamically across the landscape. The difference between these two protocols is also clear if one compares the susceptibilities measured after the same time by direct quench or in a multi-step protocol, see Fig.~\ref{fig:fig13} (a).


In Fig.~\ref{fig:fig14} we provide a real space view of the hierarchical landscape shown in Fig.~\ref{fig:fig13} (d) at $Z=10000$. To this end, we select pairs of clones belonging to the different layers of states shown in the maps, see the notations (i-iv) in Fig.~\ref{fig:fig13} (d), and we build a snapshot showing the corresponding distance fields $\Delta_{ab}^\alpha({\bf R})$.
In the snapshot we observe that the overall shade increases as the main distance between clones increases, and also that the spatial displacements become correlated over increasing lengthscales. This suggests that the hierarchical organisation of the states corresponds, in real space, to a hierarchy of length scales, presumably associated to a hierarchy of barriers between these states.

This hierarchical structure should give rise to rejuvenation and memory effects~\cite{rme1}, as discussed extensively in the spin glass literature. The various compression steps discussed above are accompanied by an aging dynamics corresponding to barriers of smaller and smaller amplitude, akin to rejuvenation effects. Given the distinct time scales mentioned above one should find the memory of the previous state by decompressing the system. This is shown in Fig.~\ref{fig:fig13} (a) (red squares). When the pressure is decreased from $Z=1000$ to $Z=600$, the susceptibility goes back to the value it had at the end of the first compression step at $Z=600$, which indicates memory. Accordingly, the heat map in Fig.~\ref{fig:fig13} (e) is similar to the one in Fig.~\ref{fig:fig13} (b), showing that the clones have retained the memory of the organisation they had at that time.

It is remarkable that such a complex organisation of the glass basin emerges in our system of $2d$ hard disks, given that no Gardner transition is expected to take place at finite pressure in this system. In the regime of pressure considered here $Z\leq10000$, the system is still pretty far from jamming, in the sense that the jamming criticality itself is not yet well-developed (for instance, the power laws characterizing the pair correlation function at contact cannot be observed at all), but our results are quite consistent with the organisation of the landscape obtained in mean-field descriptions of spin glass phases~\cite{sgbook1987}. In particular, it is striking that our heat maps resemble the pictures drawn to describe the ultrametric organisation of phase space in spin glasses~\cite{sgmap2001,sgmap2004}.

\subsection{Direct evidence for absence of Gardner transition}

\label{avoided}

\begin{figure}
\includegraphics[width=0.48\textwidth]{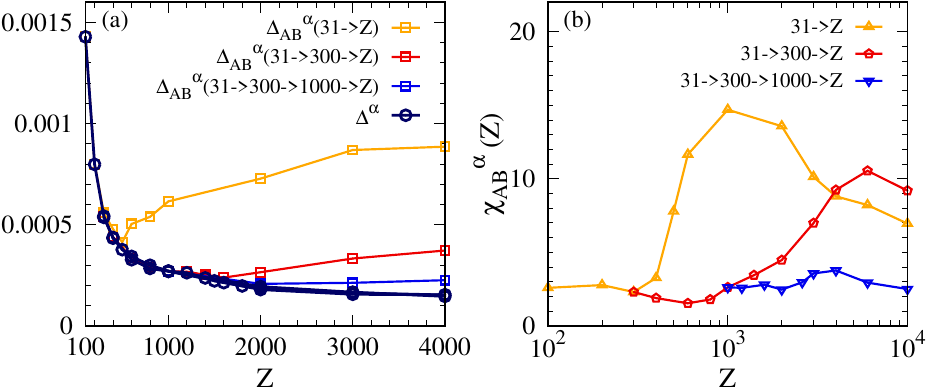}
\caption{\label{fig:fig15} Evolution of long time limits of the MSD and the average distance between clones against pressure for various protocols.
(a) The gap between $\Delta_{AB}^{\alpha}$ and $\Delta^{\alpha}$ emerges at a pressure that strongly depends on the protocol, which appears to shift the location of the Gardner crossover from $Z\approx 200$ to $Z \approx 2000$.
(b) The corresponding susceptibilities $\chi_{AB}^{\alpha}$ have a maximum that shifts to larger pressure and decreases, as the protocol is made more complicated.}
\end{figure}

The description of the glass landscape as a hierarchical organization of sub-basins suggests a strategy to maintain equilibrium up to the pressures that are above the Gardner crossover shown in Fig.~\ref{fig:fig3}. The simplest way to detect the Gardner crossover is by monitoring the difference between the long-time limits of the MSD and the averaged distance between clones. For the single sample $\alpha$ considered extensively in this work, we report those data in Fig.~\ref{fig:fig15} (a), which confirms that $Z_G \approx 200$ for this sample.

Next, we again use the idea of a multi-step protocol, but now choose carefully the intermediate pressures and duration spent at each step as follows. We first perform a quench from the initial liquid state at $Z_g=31$ up to $Z=300$. At this pressure, for the single glass sample $\alpha$, we have shown in Fig.~\ref{fig:fig8} (d) above that it takes about $t_w \approx 10^7$ for all clones to cluster inside the most stable sub-basin, which we called group$_m$. To make sure that all the clones end up in the same part of the landscape, we spend a time $t_w = 2\times10^8$ at this pressure.

Using the same set of clones, we then repeat the protocol of quenching the system to larger pressures, and we again measure the long-time limits of both $\Delta$ and $\Delta_{ab}^\alpha$ in this two-step protocol. Notably, we observe that they now start to differ from one another only when $Z$ becomes larger than $Z \approx 900$. In the language used in the previous section, it corresponds to the pressure where the underlying sub-basin of group$_m$ splits into distinct sub-basins that become dynamically inaccessible above $Z \approx 900$. This resembles a second avoided Gardner transition.

We then perform a three-step protocol $Z_g=31 \mapsto 300 \mapsto 1000$ and wait a long enough time at $Z=1000$ (we use $t_w= 2\times10^8$) so that all clones again gather into the same sub-basin at this pressure at the end of the second step. Then, we again quench further the clones to larger $Z$. The data in Fig.~\ref{fig:fig15} suggest that a third Gardner crossover now occurs at $Z \approx 2000$ in this protocol, which is about 10 times larger than the Gardner crossover detected using a direct quench.

In those three protocols, we measure the susceptibility $\chi_{ab}^\alpha$ at a fixed time $t_w = 2\times10^8$, as reported in Fig.~\ref{fig:fig15} (b). The non-monotonic behaviour observed before for a direct quench is again present in the multi-step protocols, but the location of the maximum is shifted to larger pressures. More interestingly, the amplitude of the maximum decreases as well, which agrees well with the above findings that the hierarchy of states that appear at larger pressures correspond to smaller and smaller lengthscales as well (recall the snapshots in Fig.~\ref{fig:fig14}).

In Sec.~\ref{evidence}, we obtained key signatures for the existence of a Gardner transition in our $2d$ system, with a transition possibly rounded by finite time effects. Here, we observe that by crossing the transition more slowly, these signs are shifted to another location and become less prominent. If the crossover was underlaid by a genuine phase transition, a modest shift of its location would be expected, together with sharper dynamic signatures. The results in Fig.~\ref{fig:fig15} are thus direct evidence for a transition that is avoided. In the following section, we additionally show that increasing the system size does not provide diverging correlation lengthscales either.

\section{Results for larger systems}

\label{larger}

In this final section, we expand the above analysis to larger system sizes to understand how robust the results for $N=1024$ can be. We study how the measurements related to the Gardner physics change with increasing the system size to $N=4096$ and 16384, quenching the system from $(\varphi_{g}, Z_{g}) = (0.820, 31)$ to different pressures, using $(N_{s}, N_{c}) = (100, 20)$ and maximal waiting time $t_{w} \sim 3\times10^{7}$.

\subsection{Large but finite correlation length}

\begin{figure}
\includegraphics[width=0.48\textwidth]{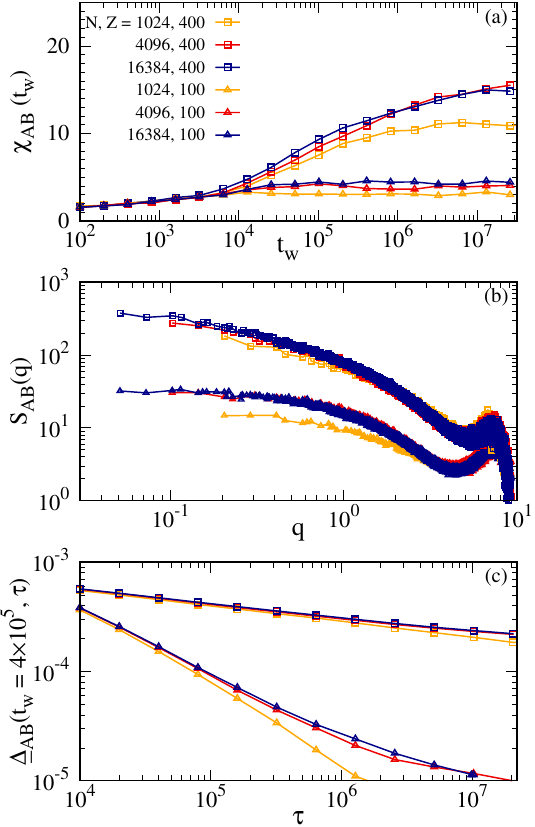}
\caption{\label{fig:fig17} Finite size effects via system sizes $N = 1024$, $4096$, and $16384$ for $Z = 100$ and $400$.
(a) Time evolution of susceptibilities $\chi_{AB}(t_{w})$.
(b) Structure factor of distance field, $S_{AB}(q)$ at $t_{w} = 2^{18}\times100$.
(c) Time decay of $\Delta_{AB}(t_{w} = 4\times10^{5}, \tau)$.
All data converge rapidly with $N$, indicating a maximal correlation length of about $L/2 \approx 30$, the linear system size for $N=4096$.}
\end{figure}

First, we present the $N$ dependence of several quantities after a quench to both $Z=100$ and $Z=400$ starting from the equilibrium system at $Z_g=31$ in Fig.~\ref{fig:fig17}, which implies that the results presented earlier for $N=1024$ do not strongly depend on the system size.

The susceptibility in Fig.~\ref{fig:fig17} together with the MSD (not shown) confirm that in large systems the dynamics is fast (no aging) and trivial (the susceptibility remains ${\cal O}(1)$) at $Z=100$. Looking at the structure factor $S_{AB}(q)$ the absence of finite size effects is also obvious (data for larger $N$ nearly superpose) and a clear plateau at low-$q$ is observed suggesting a small correlation length of a few particle diameters at most.

The situation is different at $Z=400$, where the MSD shows aging (not shown) and the susceptibility is large (of magnitude $15$ at $t_w=10^7$), and the low-$q$ part of the structure factor saturates at a much lower wavevector, indicating a much larger correlation length of the distance field. Therefore, the signatures of the Gardner crossover found in $N=1024$ are also observed in larger systems. A key observation from Fig.~\ref{fig:fig17} is the presence of a finite size effect between $N=1024$ and larger sizes, compared with the superposition of curves for $N=4096$ and $16384$. This suggests that the maximal correlation length observed in our system is of the order of the linear size of the system with $N=4096$, that is, $L/2 \approx 30$, which is consistent with the length extracted from the low-$q$ part of $S_{AB}(q)$. The presence of such a large correlation length provides the explanation why the physics related to a phase transition is obvious in our data, as it requires much larger system sizes to realise that the correlation length is actually not divergent. It would be interesting to revisit $3d$ hard sphere glasses to understand how large the correlation length can become in $d=3$.

Regarding timescales, we revisit the measurements of the distance between averaged positions, $\underline{\Delta}_{AB}(t_{w}, \tau)$ from (Eq.~(\ref{tadelta})), for different system sizes in Fig.~\ref{fig:fig17} (c). As for other quantities, we again find that the timescales extracted from these data change weakly with $N$, indicating in particular that the measured relaxation times do not appear to grow with the system size.

\subsection{Self-averaging and landscape in large systems}

The data in Fig.~\ref{fig:fig17} indicate that the maximal correlation length in the $2d$ hard disk glass after an instantaneous quench is about $\xi \approx 30$, the linear size of the system $N=4096$. Hence, a larger system with $N=16384$, corresponding to $L \approx 120$ should be understood as the superposition of several independent sub-systems. Correspondingly, the numerical measurements with $N=16384$ should be self-averaging, with the fluctuations within each glass sample, and sample-to-sample fluctuations should be much smaller.
A second consequence is that it should become less relevant to discuss the landscape of single glass samples in such large systems, because the total landscape is in fact the convolution of independent landscapes. We therefore expect that the hierarchical structure revealed when studying systems with $N=1024$ should of course still be physically relevant (for instance to account for memory and rejuvenation effects), but should also be more difficult to visualise.

\begin{figure}
\includegraphics[width=0.48\textwidth]{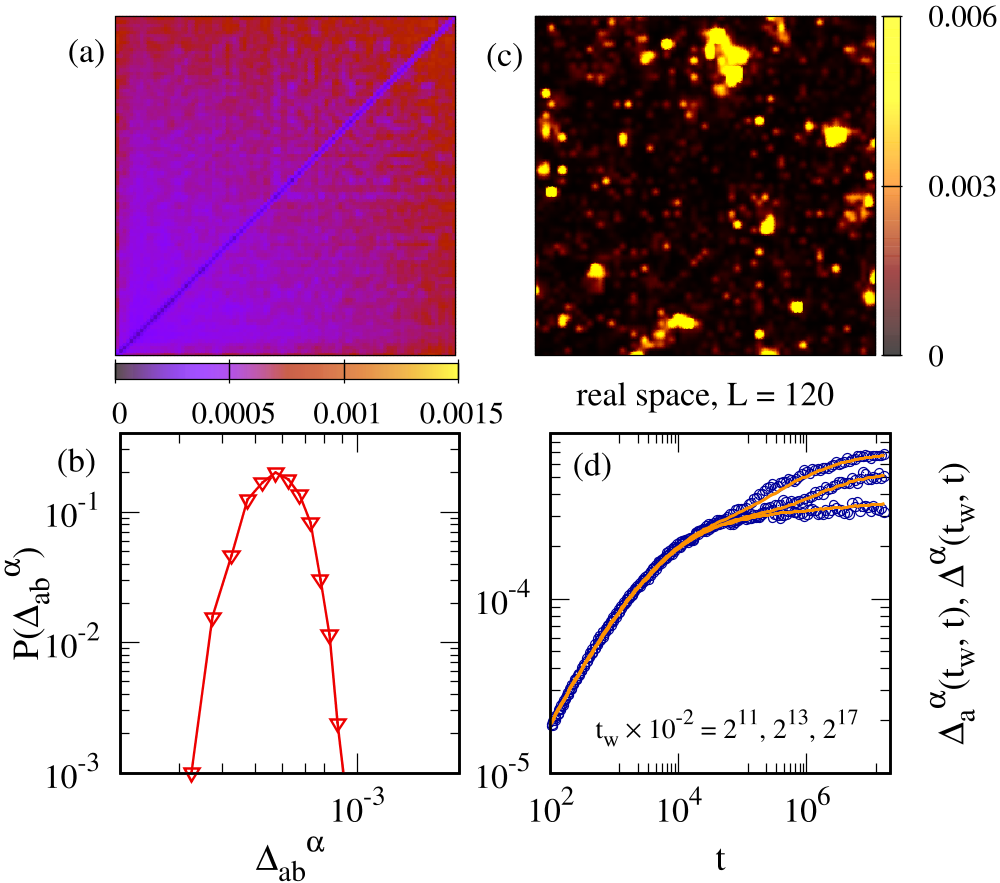}
\caption{\label{fig:fig16}
Self-averaging in a single sample $\alpha$ for $N=16384$ at $Z=600$.
(a) Heat map of $\Delta_{ab} ^{\alpha} (t_{w} = 2^{18}\times100)$, with $N_c=100$ clones.
(b) Corresponding probability distribution $P(\Delta_{ab} ^{\alpha})$.
(c) Distance field $\Delta_{ab} ^{\alpha} (\mathbf{R})$ for two randomly selected clones $a$ and $b$ at $t_{w} = 2^{18}\times100$.
(d) Aging exhibited in the MSD of a single clone $a$ (blue symbols) and in the MSD averaged over $N_c = 100$ clones (yellow lines).}
\end{figure}

To illustrate these ideas, we randomly select one glass sample with $N=16384$ and repeat the measurements presented in Fig.~\ref{fig:fig6} above.
Specifically, we quench this sample from $Z_{g} = 31$ to $Z = 600$ using $N_c=100$ clones and analyse the results after a waiting time $t_{w} = 2^{18}\times100$. The heat map of distances between clones in Fig.~\ref{fig:fig16} (a) appears featureless, which is different from the heat map constructed for the same parameters with $N=1024$ in Fig.~\ref{fig:fig6} (a). Accordingly, the distribution of distances between clones shown in Fig.~\ref{fig:fig16} (b) present no remarkable structure, in particular, it centers around a value between the two peaks of the distribution of the $N=1024$ sample in Fig.~\ref{fig:fig6} (b). When we randomly select two clones to visualise the distance field in real space, as shown in Fig.~\ref{fig:fig16} (c), one can observe large correlated domains, but the largest domains are clearly smaller than the linear size of the system, $L=120$. Therefore, the large system behaves as the superposition of many smaller parts, which explains the simple forms of the heat map and the distribution.

The self-averaging suggested by these measurements can be directly confirmed by comparing the MSD measured after a quench for a single clone (symbols), to the results obtained after averaging over $N_c$ clones (lines) in Fig.~\ref{fig:fig16} (d). Remarkably, the behaviour of one clone is similar to the averaged behaviour, and the individual events that could be clearly observed for $N=1024$ are now washed out by the self-averaging taking place inside large systems.


\section{Conclusion}

\label{conclusion}

We show that $2d$ hard disks exhibit all the phenomenology expected for systems undergoing a Gardner transition inside the glass phase~\cite{review2016}, even though no such phase transition is expected to occur in the thermodynamic limit~\cite{rg2015, rg2017, rg2018}. We observe, in particular, that hard disks display slow and aging dynamics, spatially correlated motion, non-Gaussian global fluctuations, and large sample-to-sample fluctuations, which are all reminiscent of generic observations performed in systems possessing a spin glass phase at low temperatures~\cite{sgreview2000, sg2013, sg2014, sg2018, rme1}.
In that sense, $2d$ hard disks are more similar to the $d=3$ and $d=\infty$ hard sphere models~\cite{ckpuz2014, bcjpsz2016, sz2018} than, for instance, nearly $1d$ hard disks and soft spheres for which several important indicators of a Gardner transition are absent~\cite{sbz2017, 1d2018, defect2}. We show that these observations are natural, given that a large correlation is measured as the pressure increases, estimated to be $\xi \approx 30$. Thus, we conclude that there is a sharp Gardner crossover in $2d$ hard disk glasses.

By carefully analysing the behaviour of individual glass samples, we demonstrate that the Gardner crossover is associated with the emergence of a complex free energy landscape, where the metastable glass basin breaks into smaller sub-basins with increasing pressure. We also discovered a hierarchical organisation of the landscape, associated to a hierarchy of time scales and length scales. Our study suggests that the analysis of a single glass sample with a modest system size does yield strong evidence of a phase transition~\cite{experiment2016}, but our analysis of sample-to-sample fluctuations, of a range of different system sizes and of different thermal histories has shown that no phase transition actually exists in $2d$ hard disks.

This complex organisation of the landscape contrasts strongly with the behaviour of stable glasses formed using pair potentials with no cutoff, where no jamming transition is present~\cite{sbz2017, defect2}. Therefore our findings suggest that systems approaching the jamming transition have a landscape that is more complex than ordinary amorphous solids. At the theoretical level, this complexity is signaled in the mean-field limit by the concept of a Gardner phase that is entered before reaching the jamming critical point~\cite{ckpuz2014exact}. In this approach the marginality of the Gardner phase is central to account for the properties of the jamming transition itself~\cite{ckpuz2014, jg2018}. In physical dimensions, there is no strong evidence that a Gardner phase exists in $d=3$,  and no transition is expected for $d<3$~\cite{rg2015, rg2017, rg2018}. However, our simulations in $d=2$ suggest that the phenomenology associated to a Gardner phase can be present even when a real phase transition does not exist. It would be interesting to understand better why the man-field physics is so strongly relevant in $d=2$. One possibility is that long-ranged elastic interactions between very dense hard sphere packings make the system closer to its mean-field limit.

Our results suggest, therefore, that the concepts and tools associated to the Gardner transition are not only useful to account for the jamming point, but also describe qualitatively new physical features characterizing the physical properties of a large variety of amorphous materials even away from jamming. We expect, therefore, that aging, slow dynamics, correlated particle motion, and the existence of a hierarchy of timescales, length scales and barriers to be relevant for future numerical and experimental studies of broad range of amorphous solids. We suggest, in particular, that dense amorphous materials made of simple colloidal and granular particles should display very complex dynamic features not only when approaching the glass transition but also very deep in the arrested phase.

\acknowledgments
We thank C. Artiaco, G.~Biroli, G. Parisi, C.~Scalliet, B.~Seoane, and F.~Zamponi for useful exchanges about this work.  This work is supported by a grant from the Simons Foundation (No. 454933, LB). The work of Q.L. in Montpellier was supported by the program of China Scholarships Council No.201706340021. Q. L. would like to acknowledge her supervisor N. Xu whose substantial support made the visit in Montpellier possible.

\begin{appendix}

\section{Clustering method to construct the maps}

\label{app:map}

Here we introduce the method used to sort the clones to construct the heat maps shown throughout this paper. We refer to the introduction~\cite{clustering} for advanced clustering algorithms. Our algorithm proceeds as follows. \\
\bf{Step 1.} \rm Create cluster $\mathcal{C}_{A}$ and cluster $\mathcal{C}_{B}$, place the $a$ with maximal $\sum_{b}^{N_{c}} \Delta^{\alpha}_{ab}$ as the first clone of $\mathcal{C}_{A}$ ($N_{A} = 1$), and put the rest $N_{B} = N_{c} - 1$ clones into $\mathcal{C}_{B}$. ~\\
\bf{Step 2.} \rm Transfer clone $b$ with minimal $\sum_{a\in \mathcal{C}_{A}}^{N_{A}} \Delta^{\alpha}_{ab}$ from $\mathcal{C}_{B}$ ($N_{B} = N_{B} - 1$) to $\mathcal{C}_{A}$, as the $(N_{A} = N_{A} + 1)$th clone.  ~\\
\bf{Step 3.} \rm Return to Step 2. until all clones are in cluster $\mathcal{C}_{A}$ ($N_{A} = N_{c}$). ~\\

\section{Coarse-graining method to construct the snapshot of displacement field}

\label{app:snapshot}

To compute the field $\Delta^{\alpha}_{ab}({\bf R})$, we cut the system into $N/4$ square pieces, so that the linear size of each site is $(\frac{2L}{\sqrt{N}} \times \frac{2L}{\sqrt{N}})$, and perform the averaging of Eq.~(\ref{field}) at each site,
\begin{equation}
\begin{split}
& \Delta^{\alpha}_{ab}(\mathbf{R}) = \frac{\sum_{i}^{N} \Delta^{\alpha}_{ab,i} P_{i}(\mathbf{R})}{\sum_{i}^{N}P_{i}(\mathbf{R})} \\
& P_{i}(\mathbf{R}) = \Theta(\frac{L}{\sqrt{N}} - |X - X^{\alpha}_{a,i}|) \Theta(\frac{L}{\sqrt{N}} - |Y - Y^{\alpha}_{a,i}|),
\end{split}
\end{equation}
Where $\Theta$ is step function, $(X^{\alpha}_{a,i}, Y^{\alpha}_{a,i}) = \mathbf{R}^{\alpha}_{a,i}$ is the coordinate of particle $i$ in clone $a$ and  sample $\alpha$, and $(X, Y) = \mathbf{R}$ is the center of each site.

\end{appendix}

\bibliography{reference}

\end{document}